\pdfoutput=1
\documentclass[11pt]{article}
\usepackage[margin=1in]{geometry}
\usepackage[dvipsnames]{xcolor}
\usepackage{amsmath,amssymb,amsthm}
\usepackage{mathtools}
\usepackage{graphicx}
\usepackage{url}
\usepackage[numbers,square]{natbib}
\usepackage{enumitem}
\usepackage{tikz}
\usetikzlibrary{positioning,calc,decorations.pathreplacing,fit,backgrounds,arrows.meta}
\usepackage{xspace}
\usepackage[algo2e,ruled]{algorithm2e}
\usepackage[colorlinks=true,citecolor=blue,linkcolor=blue,urlcolor=blue]{hyperref}
\usepackage{cleveref}

\usepackage{aliascnt}
\theoremstyle{plain}
\newtheorem{theorem}{Theorem}
\newaliascnt{lemma}{theorem}\newtheorem{lemma}[lemma]{Lemma}\aliascntresetthe{lemma}
\newaliascnt{proposition}{theorem}\newtheorem{proposition}[proposition]{Proposition}\aliascntresetthe{proposition}
\newaliascnt{corollary}{theorem}\newtheorem{corollary}[corollary]{Corollary}\aliascntresetthe{corollary}
\newaliascnt{observation}{theorem}\newtheorem{observation}[observation]{Observation}\aliascntresetthe{observation}
\theoremstyle{definition}
\newaliascnt{definition}{theorem}\newtheorem{definition}[definition]{Definition}\aliascntresetthe{definition}
\newaliascnt{remark}{theorem}\newtheorem{remark}[remark]{Remark}\aliascntresetthe{remark}

\crefname{section}{Section}{Sections}        \Crefname{section}{Section}{Sections}
\crefname{subsection}{Section}{Sections}     \Crefname{subsection}{Section}{Sections}
\crefname{figure}{Figure}{Figures}           \Crefname{figure}{Figure}{Figures}
\crefname{algocf}{Algorithm}{Algorithms}     \Crefname{algocf}{Algorithm}{Algorithms}
\crefname{equation}{}{}                      \Crefname{equation}{}{}
\crefname{theorem}{Theorem}{Theorems}       \Crefname{theorem}{Theorem}{Theorems}
\crefname{lemma}{Lemma}{Lemmas}             \Crefname{lemma}{Lemma}{Lemmas}
\crefname{proposition}{Proposition}{Propositions} \Crefname{proposition}{Proposition}{Propositions}
\crefname{corollary}{Corollary}{Corollaries} \Crefname{corollary}{Corollary}{Corollaries}
\crefname{observation}{Observation}{Observations} \Crefname{observation}{Observation}{Observations}
\crefname{definition}{Definition}{Definitions} \Crefname{definition}{Definition}{Definitions}
\crefname{remark}{Remark}{Remarks}          \Crefname{remark}{Remark}{Remarks}

\DeclareRobustCommand{\opt}{\mathrm{OPT}}
\newcommand{\optC}{\opt_{\mathrm{cen}}}      %
\newcommand{\optR}{\opt_{\mathrm{msr}}}      %
\DeclareRobustCommand{\eps}{\varepsilon}
\newcommand{\Ball}{B}
\newcommand{\wt}[1]{\widetilde{#1}}
\DeclareRobustCommand{\median}{\mathsf{med}}
\newcommand{\MedDist}{\mathrm{md}}

\newcommand{\poly}{\mathrm{poly}}
\newcommand{\polylog}{\mathrm{polylog}}
\newcommand{\Test}{\textsc{Cover}\xspace}
\newcommand{\Peel}{\textsc{Peel}\xspace}
\newcommand{\Sketch}{\textsc{Sketch}\xspace}
\newcommand{\safe}{\textsc{full}\xspace}
\newcommand{\uns}{\textsc{partial}\xspace}
\newcommand{\SO}{\mathsf{SO}}
\newcommand{\whp}{w.h.p.\xspace}

  \newcommand{\Good}{\mathcal E}          %
  \definecolor{clA}{RGB}{31,110,180}
  \definecolor{clB}{RGB}{214,96,40}
  \definecolor{okgreen}{RGB}{35,132,67}
  \definecolor{badred}{RGB}{200,40,40}

\title{$k$-Center under Weak and Strong Distance Oracles with an Application to Minimum Sum of Radii Clustering}
\author{Zachary Friggstad\thanks{University of Alberta. \texttt{zacharyf@ualberta.ca}}
\and Mahya Jamshidian\thanks{University of Alberta. \texttt{mjamshidian@ualberta.ca}}
\and Ali Vakilian\thanks{Rutgers University. \texttt{a.vakilian@rutgers.edu}}}
\date{}

\begin{document}
\maketitle

\begin{abstract}
In the \emph{weak--strong distance oracle} model of Bateni, Dharangutte, Jayaram and Wang (COLT 2024), a metric $(X,d)$ on $n$ points is accessed through two oracles: a cheap \emph{weak} oracle that returns each queried distance correctly with probability $1-\delta$ and permits arbitrary values on corrupted pairs, chosen by an adversary who knows the entire metric and all corrupted pairs, and an expensive \emph{strong} oracle that reveals a point exactly.
The goal is to cluster well while making as few strong queries as possible.
This is a query-efficient unsupervised learning problem: reliable information is costly, while errors in the cheap observations persist across repeated queries.

Our main result is a $(6+\eps)$-approximation for $k$-center using $O(k\log n\log(\eps^{-1}\log n))$ strong point queries, for any fixed $\eps>0$.
This improves both the $(14+\eps)$ approximation factor and the $O(k\log^2 n\log(\eps^{-1}\log n))$ query bound of Bateni et al., and our algorithm is simpler: it makes one greedy pass per radius guess instead of $O(\log n)$ rounds of random sampling. It also improves the query complexity of the recent $(6+\eps)$-approximation of Pradhan, Bhattacharya and Jaiswal (AISTATS 2026), in the variant of the model they study.

Our algorithm outputs an explicitly known  \emph{sketch} of an otherwise unknown metric.
The constant-factor $k$-center guarantee is the least of what this yields; the useful part is that a downstream algorithm is handed a concrete metric on at most $k$ points that it can compute with directly, with no further queries and no noise.
Specifically, we show an application of such sketch for \textsc{Minimum Sum of Radii} (MSR), the objective left unexplored in this weak--strong oracles model and the one where the absence of a single relevant scale makes cheap queries least useful.
Given the sketch there is a simple FPT algorithm that returns a constant-factor solution of MSR using $2^{O(k\log k)}\cdot O(\log n\log\log n)$ strong queries.

We complement the upper bound with an $\Omega\bigl(k\log(n/k)\bigr)$ lower bound for $k$-center with assignment that holds for \emph{row-local} algorithms.
These algorithms assign two unqueried points to the same center whenever their weak distances to the strong-queried points are exactly the same.
Our algorithm and those of Bateni et al.\ and Pradhan et al.\ are all row-local.
Our query bound is therefore optimal for this class up to the $O(\log\log n)$ factor.

\end{abstract}

\section{Introduction}
\label{sec:intro}

Clustering algorithms are usually designed under the assumption
that pairwise distances can be obtained at no cost.
From the perspective of algorithmic learning theory,
this assumption hides an information-acquisition cost:
in many unsupervised tasks, the learner must decide which
distances are worth obtaining reliably and which can be
estimated from cheaper, imperfect feedback.
This assumption is a poor fit for a growing class of modern applications, where the ``points'' are documents, images, or user profiles and the metric is not tabulated in advance but computed on demand by a learned similarity model.
Computing one such distance is expensive, and a typical pipeline hedges by keeping two models on hand: a large, accurate one that is costly to invoke, and a small, cheap one that is fast but occasionally badly wrong~\cite{bateni2024metric}.
One would like to spend the expensive model sparingly and let the cheap one do the bulk of the work. 

Bateni, Dharangutte, Jayaram and Wang~\cite{bateni2024metric} formalized this trade-off as the \emph{weak--strong distance oracle} model.
A weak query on a pair $\{x,y\}$ returns the true distance $d(x,y)$ with probability $1-\delta$ for some constant $\delta \in (0,1/2)$ and an arbitrary value otherwise. Corruptible pairs are chosen independently, but an adversary may select their reported values \emph{after seeing the entire set of corruptible pairs and the true distances of all pairs}. Crucially, the answer for a given pair is fixed once and for all, so repeating a query does not resample the noise; i.e., the noise is persistent.
A strong query on a single point $x$ returns an object $\SO(x)$, think of it as an exact embedding, and any two such objects together reveal the exact distance between their points.
Weak queries are free; strong queries are the resource to be minimized.
This makes strong-query complexity an analogue of label or verification complexity in active and weak-supervision learning, while the output remains a clustering of an unknown metric.
In this model, Bateni et al.~\cite{bateni2024metric} obtained constant-factor approximations for $k$-center with $O(k\log^2 n\log\log n)$ strong point queries, and for $k$-median and $k$-means with $O(k\log^2 n)$. Moreover, they also showed that $\Omega(k)$ strong queries are necessary for any bounded approximation.

We revisit $k$-center. Recall that \emph{$k$-center} asks for at most $k$ points of $X$ (the \emph{centers}) minimizing the largest distance from any point to its nearest center; we write $\optC$ for this optimal radius.
The algorithm of Bateni et al.~\cite{bateni2024metric} is a recursive ``sample-and-cover'' scheme.
They guess a radius $R\approx2\optC$ by binary search.
Each round samples $O(k\log n)$ points and strong-queries them.
Using these points as witnesses and a median of weak queries, it estimates which of the remaining points fall into an optimal ball, and assigns those points.
Each round covers a constant fraction of the surviving points, so $O(\log n)$ rounds suffice to cover everything, which is where the second logarithmic factor in their query bound comes from.
Pradhan, Bhattacharya and Jaiswal~\cite{pradhan2026improved} subsequently improved on Bateni et al.'s $k$-center approximation by a different, \emph{ball-carving} approach, obtaining a $(6+\eps)$-approximation with $O(k^3\log^2 n\log(\eps^{-1}\log n))$ strong \emph{distance (edge)} queries. This is a result for the edge-query model, where a strong query reveals one pairwise distance, whereas our result concerns the point-query model, where the distances between strongly queried points can be recovered exactly. Because the two results count different query primitives, their query bounds are not directly comparable.

Our starting point is the natural greedy procedure for $k$-center in this model, which is simpler than that of~\cite{bateni2024metric}.
It is a single Gonzalez-style~\cite{gonzalez1985clustering} pass over the points, with no recursion on residual point sets.
It opens a new center only when a point fails a deterministic ``test''.
Relative to the strong-point result of Bateni et al., this gives an improvement in both approximation and query complexity: we obtain a $(6+\eps)$-approximation for fixed $\eps>0$ using $O(k\log n\log(\eps^{-1}\log n))$ strong-point queries, improving their $(14+\eps)$ approximation and $O(k\log^2 n\log(\eps^{-1}\log n))$ strong-point query bound.
Moreover, the same algorithm can be implemented using $O(k^2\log n\log(\eps^{-1} \log n))$ strong \emph{edge} queries, or $O(k^2\log n\log\log n)$ for fixed $\eps$.\footnote{Throughout the remainder of the paper, we treat $\eps$ as a constant and omit the $\log(\eps^{-1})$ dependence from the query complexity, since both our results and all prior results incur the same dependence on $\eps$. This factor arises solely from the binary search over a geometric grid with ratio $1+\eps$ (\Cref{lem:search}).} Indeed, each of its $O(k\log n)$ exact-routing steps per radius guess requires distances only to the at most $k$ current representatives, so it incurs at most $k$ edge queries. Thus our $(6+\eps)$ approximation also improves the edge-query bound of Pradhan et al.~\cite{pradhan2026improved} by a factor of $\Omega(k\log n)$; see \Cref{rem:edge} for the formal accounting.

It is also, in a sense we make precise, the best one can do without changing strategy.
Our third contribution is a lower bound showing that $\Omega\bigl(k\log(n/k)\bigr)$ strong queries are necessary for any algorithm that shares one structural feature with ours: two points that are never strong-queried and have exactly the same weak distances to the strong-queried points are assigned to the same center.
We call such algorithms \emph{row-local}; our procedure and those of~\cite{bateni2024metric} and~\cite{pradhan2026improved} are all of this kind.
So the second logarithm we save is real, but the first one is not removable by any algorithm of this kind. 

\smallskip
As an application, and our second contribution, we use the sketch to give the first approximation algorithm for {\em Minimum Sum of Radii} (MSR) in the weak--strong model; we believe the reduction, which turns a metric-clustering task in this model into a branching search over the $O(k)$ sketch points, may be of independent interest.
MSR requires one to select at most $k$ balls, centered at points of $X$, whose union covers $X$, so as to minimize the \emph{sum} of the radii.
The objective sits between $k$-center (minimize the maximum radius) and $k$-median (minimize the sum of assignment distances).
MSR is a demanding test, for two reasons.
First, it has not been studied in the weak--strong model at all.
Second, it is hard precisely because the $k$ radii of an optimal solution may differ wildly, so there is no single scale at which to discretize the instance.
This is what sets it apart from $k$-center, whose algorithm in this model~\cite{bateni2024metric} fixes a single radius $R\approx2\optC$ and only has to search for it.
For MSR, no single radius can stand in for the whole solution.

\subsection{Our results}
\label{sec:results}

Throughout, $\optR(X,k)$ denotes the optimal MSR cost with $k$ balls and $\optC(X,k)$ the optimal $k$-center radius; ``\whp'' means with probability $1-n^{-\Omega(1)}$.
We assume, as is standard in this model~\cite{bateni2024metric}, that the aspect ratio of $(X,d)$ is $\poly(n)$. Our main result is the sketch guarantee, which we prove as \Cref{lem:sketch}.

\begin{theorem}
\label{thm:intro-sketch}
For every fixed $\eps>0$, there is an algorithm that runs in polynomial time, uses $O(k\log n\log\log n)$ strong point queries and, \whp, outputs a partition of $X$ into $m\le k$ clusters $X_1,\dots,X_m$ with representatives $\rho_1,\dots,\rho_m$ such that
\begin{enumerate}[label=(\roman*),topsep=2pt,itemsep=1pt]
  \item every $\rho_i$ has been strong-queried, and \emph{all} their pairwise distances are known; and
  \item  $d(x,\rho_i)\le (6+\eps)\,\optC(X,k)$ for every $x\in X_i$ and every $i$.
\end{enumerate}
\end{theorem}
We emphasize that the \whp is only over the random perturbations in the weak oracle.
Our algorithm is deterministic otherwise.

The bound in (ii) is a $k$-center approximation, which was already known in this model~\cite{bateni2024metric}; what is new is that we obtain it more cheaply and more simply.
Property (i) is what lets the application in \Cref{sec:fptmsr} treat the output as an ordinary metric on $\le k$ points: after the strong queries are spent, the sketch is exact and the weak oracle plays no further role.
We do not read (i) as a separate feature, any $k$-center routine that queries its centers has it.
But it is the property our MSR algorithm relies on, so we state it explicitly.
In particular, the representatives themselves are a $k$-center solution, so we get the following.
\begin{corollary} \label{cor:intro-kcenter}
    For every fixed $\eps>0$ there is an algorithm that, using $O(k\log n\log\log n)$ strong point queries, outputs \whp a $(6+\eps)$-approximate $k$-center solution.
\end{corollary}
A second feature is worth flagging.
The feasibility test underlying \Cref{thm:intro-sketch} has one-sided error (\Cref{thm:cover}): when it reports that $k$ balls of radius $\tau$ do not cover $X$, that verdict is a genuine certificate of infeasibility.

\smallskip
Our application is the following MSR result, which we prove as \Cref{thm:fptmsr}.

\begin{theorem}
\label{thm:intro-msr}
There is an algorithm that, \whp, outputs an MSR solution with at most $k$ balls and total cost at most $40\cdot\optR(X,k)$.
It uses $f(k)\cdot O(k\log n\log\log n)$ strong point queries and runs in $f(k)\cdot\poly(n)$ time, where $f(k)=2^{O(k\log k)}$.
\end{theorem}

A linear dependence on $k$ is known to be necessary: no algorithm outputs a bounded approximation with $o(k)$ strong point queries (\Cref{thm:lb}, adapting the $\Omega(k)$ bound of~\cite{bateni2024metric}). Our third result says that for $k$-center the true price is a logarithmic factor higher, for every algorithm built the way ours is.
Call an algorithm \emph{row-local} (\Cref{def:rowlocal}) if it assigns any two points that are neither strong-queried nor centers to the same center whenever their weak distances to the strong-queried points are exactly the same; the queried points and the centers may still be chosen using all weak answers.

We prove the following lower bound as \Cref{cor:rowlocal}.

\begin{theorem}
\label{thm:intro-rowlocal}
Let $\alpha\ge1$ and $c>0$ be constants, and let $k\le n^{1-c}$ be even and divide $n$.
Every row-local algorithm that outputs an $\alpha$-approximate $k$-center solution \emph{together with its assignment}, with probability more than $1/4$, makes $\Omega\bigl(k\log(n/k)\bigr)$ strong point queries.
\end{theorem}

Since \Cref{thm:intro-sketch} outputs a partition, it is a solution with assignment, and a mild re-routing makes it row-local (\Cref{prop:ours-rowlocal}); the two bounds therefore meet, up to $O(\log\log n)$, for every $k\le n^{1-c}$.

However, for MSR a large gap remains: our upper bound carries the fixed-parameter factor $f(k)=2^{O(k\log k)}$, and whether it can be avoided, i.e., whether a $\poly(k)\cdot\polylog(n)$-query algorithm exists, is open.
We have made no attempt to optimize the constant $40$; in the classical setting the best known polynomial-time factor is $3+\eps$~\cite{buchem2024} and the best FPT factor is now $1+\eps$~\cite{fptmsr}, which followed earlier FPT constant-factor approximations~\cite{bandyapadhyay2023fpt,chen2024parameterized}.
In the next section, we comment how a $(1+\varepsilon)$-approximation for MSR in the weak-strong oracle model is not possible using $o(n)$ strong oracle queries.

\subsection{Technical overview}
\label{sec:overview}

\paragraph{Buying a sketch with $O(k\log n)$ strong queries.}
Fix a guess $\tau$ for $\optC(X,k)$ and process the points in arbitrary order, maintaining clusters with representatives.
The naive procedure, which strong-queries every point and attaches it to a representative within $2\tau$, opening a new cluster if none exists, is a Gonzalez-style greedy algorithm~\cite{gonzalez1985clustering} and uses $n$ strong queries.
To do better we must route most points using weak queries only, and a single weak query is useless because it is wrong with constant probability.

The fix is \emph{median amplification against a witness set}, which is also a building component in the approach of~\cite{bateni2024metric}.
Once a cluster has accumulated $L=\Theta(\log n)$ points, we freeze $L$ of them as a witness set $W(c)$ and call the cluster \safe. Each witness has already been \emph{strong-queried}; in particular, the witness set is formed entirely before any point is admitted to that cluster by a weak median test.
Every witness is within $2\tau$ of $c$, by construction.
A later point $x$ can then be tested against $c$ using $L$ weak queries, one per witness.
We take the median of these $L$ answers.
More than half of them are uncorrupted \whp, so the median lies between the smallest and largest \emph{true} distances $d(x,q)$ over the witnesses $q$.
Each such $d(x,q)$ is within $2\tau$ of $d(x,c)$ by the triangle inequality, so the median has value in $d(x,c)\pm2\tau$.
So $\Theta(\log n)$ weak queries simulate one strong query, up to additive $2\tau$, which is all we need at resolution $\tau$.
Only points that fail this test for every \safe representative are strong-queried, and a short charging argument (\Cref{lem:queries}) shows \whp there are at most $kL +1$ of them.

\paragraph{The application to MSR: peeling the largest ball.}
Let $\mathcal O=\{\Ball(o_i,r_i)\}_{i\le k}$ be an optimal MSR solution and let $\Ball(o^\star,r_{\max})$ be its largest ball.
Because $\tau \coloneqq \optC(X,k)\le r_{\max}$, the sketch of \Cref{thm:intro-sketch} has resolution $O(r_{\max})$, fine enough to locate $\Ball(o^\star,r_{\max})$ to within a constant factor of its own radius, and this is the whole point of running the sketch afresh at every level.

Let $\rho$ be the representative of the cluster containing a point covered by $\Ball(o^\star,r_{\max})$, so $d(\rho,o^\star)\le r_{\max}+O(\tau)$.
If we knew $\rho$ and a guess $R\in[2r_{\max},4r_{\max}]$, we could delete $Y(\rho,R) \coloneqq \bigcup\bigl\{X_{\rho'}: d(\rho,\rho')\le R+O(\tau)\bigr\}$, a set defined \emph{purely by distances between representatives}, all of which are known exactly.
Two applications of the triangle inequality (\Cref{lem:sandwich}) sandwich it:
\begin{align*}
    \Ball(o^\star,r_{\max})\ \subseteq\ Y(\rho,R)\ \subseteq\ \Ball(\rho,\,R+O(\tau)).
\end{align*}
The left inclusion says deleting $Y$ removes an entire optimal ball, so the residual optimum drops by $r_{\max}$; the right says the single ball we add to pay for it has radius $O(r_{\max})$.
Summing over levels telescopes to $O(\optR)$.
We do not know $\rho$ or $R$, so we branch over both: $\le k$ choices of representative and $O(\log k)$ of radius per level, whence $f(k)=k!\,(\log k)^{O(k)}$.
We stress how little this argument needs.
There is no case analysis, no appeal to the weak oracle after the sketch is built, and no structure beyond the triangle inequality.

We note that similar ideas have been used in more complex ways to design FPT algorithms for MSR when the metric is explicitly known, e.g. \cite{bandyapadhyay2023fpt}, \cite{chen2024parameterized} and ultimately culminating in a $(1+\eps)$-approximation in FPT time \cite{fptmsr}.
However, a $(1+\eps)$-approximation is not possible in the weak-strong distance oracle model even when $k = 1$ without using $\Omega(n)$ strong queries.
For example, consider the following two metrics over the same set of points $V \cup \{w\}$.
The first metric has $d(u,w) = 1$ for $u \in V$ and $d(u,v) = 2$ for distinct $u,v \in V$.
The second metric has the same distances except $d(v',w) = 2$ for a single unknown $v' \in V$.
Without strong querying $\Omega(n)$ points we would not be able to identify $v'$, in which case we could not know if the optimum MSR solution cost was $1$ or $2$. 

\paragraph{The lower bound.}
The hard instance consists of $k/2$ pairs of \emph{twin} clusters of size $n/k$.
The algorithm knows the union of every two twins, but not how it is split into the two clusters.
Distances are $\eps$ inside a cluster and $1$ otherwise.
If a weak answer between two twins is corrupted, the adversary always reports the wrong value.
Consider a pair with only a few queried points.
Take a point $x$ of one twin whose weak answers to these queried points are \emph{all} corrupted.
Take a point $y$ of the other twin whose weak answers to them are all correct.
Then $x$ and $y$ have exactly the same weak distances to the queried points (\Cref{lem:collision}).
So a row-local algorithm assigns $x$ and $y$ to the same center.
This center is wrong for one of them, so the solution has cost $1$, while the optimum is $\eps$ (\Cref{lem:confusion}).

The counting is then routine.
Suppose a pair has at most $\ell=\lfloor\frac12\log_{1/\delta}(n/k)\rfloor$ queried points.
Then one twin has $\Omega(\sqrt{n/k})$ points like $x$, and the other twin has $\Omega(\sqrt{n/k})$ points like $y$ (\Cref{lem:lb-counting}).
So these points cannot all be centers.
With at most $k\ell/4$ queries, more than a quarter of the $k/2$ pairs have at most $\ell$ queried points.
One of these pairs gives such $x$ and $y$ that are not centers.

\subsection{Related work}
\label{sec:related}

\paragraph{The weak--strong oracle model.}
Bateni et al.~\cite{bateni2024metric} introduced the model and gave constant-factor algorithms using $\widetilde O(k)$ strong queries for $k$-center, $k$-median and $k$-means, a $\widetilde\Theta(k^2)$ characterization for the edge-query variant, and an $O(\sqrt{\log n})$-approximation for MST using \emph{no} strong queries.
Pradhan et al.~\cite{pradhan2026improved} studied $k$-center and $k$-means in the edge-query variant, improving the approximation factor of Bateni et al.\ for $k$-center to $6(1+\eps)$ and their query bound for $k$-means to $O(k^2\log^2 n)$.
Fair clustering in the same model was studied by Braverman et al.~\cite{braverman2025fair}.

The model is related to a long line of work on clustering from noisy or limited supervision, e.g.\ same-cluster queries~\cite{ashtiani2016clustering} and clustering with noisy queries~\cite{mazumdar2017clustering}.
The distinguishing feature of~\cite{bateni2024metric} is that the noise is \emph{persistent}: re-querying a corrupted pair returns the same corrupted answer.
This rules out naive repetition and motivates aggregation across distinct witness pairs. Persistence is also what makes our lower bound possible: a point whose links into the queried set happen to be entirely corrupted stays corrupted, and no amount of re-querying recovers it.
In \Cref{sec:lb-barrier} we show that the row-local restriction cannot be dropped: on the instance of \Cref{sec:lb-instance}, comparing unqueried points with each other recovers the clusters with no strong queries.

\paragraph{Other noisy-oracle and weak/strong information models.}
Galhotra, Raychaudhury and Sintos~\cite{galhotra2024clustering} and Raychaudhury et al.~\cite{raychaudhury2025metric,raychaudhury2026weak} study clustering using \emph{quadruplet comparisons}, which report relative orderings of two distances, rather than the corrupted numerical distances considered here. Their algorithms and query bounds therefore address distinct information primitives, even though both lines of work ask how to substitute weak feedback for expensive exact distances. Related perspectives include active learning with weak and strong labelers~\cite{zhang2015active}, noisy comparisons without resampling~\cite{braverman2008noisy}, and clustering with cheap-weak and expensive-strong signals~\cite{silwal2023kwikbucks}; these results provide context, not reductions between our oracle models.

\paragraph{Minimum sum of radii.}
MSR was introduced by Doddi et al.~\cite{doddi2000approximation} and is NP-hard even in metrics of constant doubling dimension~\cite{gibson2010metric}.
Charikar and Panigrahy~\cite{charikar2004clustering} gave the first constant factor, $3.504$, via a primal--dual algorithm; this stood until Friggstad and Jamshidian~\cite{friggstad2022improved} improved it to $3.389$, and Buchem et al.~\cite{buchem2024} to $3+\eps$, also handling outliers.
Exact algorithms are known in $n^{O(\log n\log\Delta)}$ time~\cite{gibson2010metric} where $\Delta$ is the aspect ratio of the metric  and in special metrics~\cite{gibson2012clustering,behsaz2015minimum}; a dynamic version is studied in~\cite{henzinger2020dynamic}.

Parameterizing by $k$ has been very effective for MSR and its constrained variants.
Inamdar and Varadarajan~\cite{inamdar2020capacitated} gave the first FPT constant factor for the capacitated version, since improved by Jaiswal, Kumar and Yadav~\cite{jaiswal2024fpt}.
Approximation algorithms with a factor $(2+\eps)$ for MSR were obtained independently by Bandyapadhyay, Lochet and Saurabh~\cite{bandyapadhyay2023fpt} and by Chen et al.~\cite{chen2024parameterized}, both in time $2^{O(k\log(k/\eps))}\poly(n)$.
Recently, an FPT approximation \emph{scheme} was given for any metric \cite{fptmsr}, extending previous work on FPT schemes for metrics of bounded doubling dimension~~\cite{banerjee2025improved}.
Our contribution is not a better factor, but the observation that an FPT branching scheme can be driven entirely by an $O(k)$-size sketch, and so transfers to a setting where the metric itself is unavailable.

\paragraph{Organization.}
\Cref{sec:prelim} fixes the model and notation. \Cref{sec:kcenter} contains the main result: the feasibility test (\Cref{sec:cover}), the single lemma we need about the weak oracle together with the correctness and the query bound of the test (\Cref{sec:median}), and finally the sketch guarantee of \Cref{lem:sketch}. 
\Cref{sec:rowlocal} proves the $\Omega\bigl(k\log(n/k)\bigr)$ lower bound for row-local algorithms and shows, in \Cref{sec:lb-barrier}, that the restriction is needed.
\Cref{sec:fptmsr} gives the application to MSR, preceded by the $\Omega(k)$ lower bound that calibrates it.

\section{Preliminaries}
\label{sec:prelim}

\paragraph{The weak--strong distance-oracle model.}
We work in the \emph{weak--strong distance-oracle model} of Bateni et al.~\cite{bateni2024metric}.
The input is a finite metric space $(X,d)$ on $n=|X|$ points, where $d:X\times X\to\mathbb R_{\ge0}$ is \emph{not} given explicitly.
The algorithm accesses distances through two oracles.
\begin{itemize}[leftmargin=*,itemsep=5pt]
  \item \textbf{Weak oracle.}
        For a pair $x\ne y$, a weak query returns a value $\wt d(x,y)$ with $\wt d(x,y) = d(x,y)$ with probability $1-\delta$
        for a fixed constant $\delta\in(0,1/2)$. 
        More precisely, an independent Bernoulli($\delta$) variable marks each unordered pair as \emph{corruptible}; let $E_{\mathrm{bad}}$ be the resulting random set.
        After \emph{all} of $E_{\mathrm{bad}}$ has been sampled, an adversary who knows both the \emph{entire set $E_{\mathrm{bad}}$} and the \emph{true distances $d(u,v)$ of every pair} $\{u,v\}\subseteq X$ may choose the reported values $\wt d(x,y)$ on $E_{\mathrm{bad}}$ arbitrarily and jointly.
        Outside $E_{\mathrm{bad}}$ the weak answers equal their true distances.
        The randomness is independent across distinct pairs and is drawn \emph{once} per
        pair, so repeated weak queries on the same pair return the same value.
        Thus only the corruption \emph{locations} are independent; the adversarially chosen corrupted \emph{values} need not be independent, bounded, or oblivious to the metric or corruption pattern. Corrupted values need not satisfy the triangle inequality, nor do the weak answers as a whole have to define a metric; only the underlying $d$ is assumed metric.
        Weak queries are free and we make no attempt to minimize their number.

  \item \textbf{Strong point oracle.}
        For a point $x$, a strong query returns an object $\SO(x)$. A single $\SO(x)$
        reveals no distance, but from $\SO(x)$ and $\SO(y)$ one recovers $d(x,y)$ exactly. Strong queries are the resource to be minimized.
\end{itemize}

The following immediate consequence of the strong oracle is used throughout.
Once a set $S\subseteq X$ has been strong-queried, \emph{all} pairwise distances inside $S$ are known exactly and at no further cost.

\begin{remark}
\label{rem:edge}
Bateni et al.~\cite{bateni2024metric} consider two forms of strong oracle: the \emph{point} queries above, and \emph{edge} queries, which return $d(x,y)$ for a single requested pair.
This paper works throughout with point queries.
The two models are polynomially related.
The information obtained from $q$ point queries is exactly a set of $\binom q2$ pairwise distances, so it can be reproduced by $\binom q2$ edge queries.
In the other direction, one edge query is simulated by two point queries.
Our bounds therefore transfer with a quadratic loss; \Cref{thm:intro-sketch} becomes $\widetilde O(k^2)$ edge queries, which is the correct order, since \cite{bateni2024metric} proves matching $\widetilde\Theta(k^2)$ upper and lower bounds in the edge-query model. For our $k$-center algorithm, however, directly implementing \Test with edge queries is substantially cheaper than retrieving all pairs among the strong-queried points. In step~(B) of \Cref{alg:cover}, a point $x$ needs exact distances only to the at most $m$ existing representatives, at a cost of at most $m$ edge queries. This suffices for the witness invariant: every witness is certified to be within $2\tau$ of its representative, and the median tests require no exact distances between witnesses. By \Cref{lem:queries}, with high probability at most $mL+1$ points reach step~(B) in each execution, giving at most $m(mL+1)$ edge queries per execution. Summing over the $O(\log(\eps^{-1}\log n))$ executions in \Cref{lem:search}, with $m=k$ and $L=\Theta(\log n)$, gives a $(6+\eps)$-approximation using $O(k^2\log n\log(\eps^{-1}\log n))$ strong edge queries, or $O(k^2\log n\log\log n)$ for fixed $\eps$. This improves the $O(k^3\log^2 n\log(\eps^{-1}\log n))$ edge-query bound of~\cite{pradhan2026improved}. 
\end{remark}

Throughout the paper, we assume the {\em aspect ratio} $\Delta\coloneqq\max_{x\ne y}d(x,y)/\min_{x\ne y}d(x,y)$ of the point set $(X,d)$ satisfies $\Delta=\poly(n)$, and a value $D$ with $\min_{x\ne y}d(x,y)\le D\le\max_{x\ne y}d(x,y)$ is known.
This is the standard assumption in this model~\cite{bateni2024metric}; it is what makes the radius search described below cost $O(\log\log n)$. %

\paragraph{Clustering objectives.}
We consider two well-studied clustering objectives in this paper: $k$-center and minimum sum-of-radii~(MSR).

\smallskip
\emph{$k$-center.}
Given a metric space $(X, d)$, and a target number of clusters $k$, the goal is to choose at most $k$ centers from $X$ minimizing the largest distance from a point of $X$ to its nearest center.
Formally, $\optC(X,k) \coloneqq \min_{C\subseteq X, |C|\le k} \max_{x\in X} d(x,C)$.

\smallskip
\emph{Minimum sum of radii.}
An MSR solution for $(X,k)$ is a set of at most $k$ balls $\Ball(c_i,r_i)$ with $c_i\in X$ whose union covers $X$; its cost is $\sum_i r_i$.
We write $\optR(X,k)$ for the minimum cost.

Throughout, $(X,k)$ refers to the original instance.
For a set $U\subseteq X$ and a budget $m\le k$ we write $\optC(U,m)$ and $\optR(U,m)$ for the same objectives on $U$, with centers still chosen from all of $X$; the algorithm of \Cref{sec:fptmsr} recurses on such sub-instances.
After a part of $U$ is deleted, the surviving balls of an optimal solution keep their centers, so residual optima only decrease.
We record the inequalities linking the two objectives; the same proof gives them for every sub-instance $(U,m)$.

Let $\{\Ball(o_i,r_i)\}_{i\le k}$ be an optimal MSR solution for $(X,k)$ and put $r_{\max}\coloneqq\max_i r_i$.
Hence,
\begin{equation}
\label{eq:cen-le-rmax}
  \optC(X, k) \le r_{\max} \le \optR(X, k) \le k \cdot \optC(X, k) .
\end{equation}

Our guarantees are randomized, and the results hold \emph{with high probability} (\whp) which means with probability at least $1-n^{-\Omega(1)}$.
The analysis uses the multiplicative Chernoff bound in the form: if $Z\sim\mathrm{Bin}(L,p)$ with $p>1/2$, then $\Pr[Z\le L/2]\le e^{-2(p-1/2)^2L}$.

\section{Main result: $k$-center and a sketch of the metric}
\label{sec:kcenter}

This section proves our main result, \Cref{thm:intro-sketch}.
Fix a working set $U\subseteq X$, a budget $m\le k$, a threshold $\tau>0$, and a \emph{witness size} $L = \frac{3}{(1/2 - \delta)^2} \cdot \log n$.

\subsection{The procedure}
\label{sec:cover}

The procedure $\Test(U,m,\tau,L)$ processes the points of $U$ in an arbitrary order, maintaining a list $C$ of representatives, each owning a cluster.
A cluster is \uns while it has fewer than $L$ points and \safe once it reaches $L$; at the moment it turns \safe we freeze its $L$ current members as a \emph{witness set} $W(c)$.
Importantly, \emph{every member of $W(c)$ has already been strong-queried} and verified to be at distance at most $2\tau$ from $c$: weak routing begins for a cluster only after this certificate exists (formalized in \Cref{lem:witness}).
For a point $x$ and a \safe representative $c$ we write
\[
  \MedDist(x,c)\ :=\ \median\bigl(\{\wt d(x,q):q\in W(c)\}\bigr),
\]
the median of $L$ weak queries.

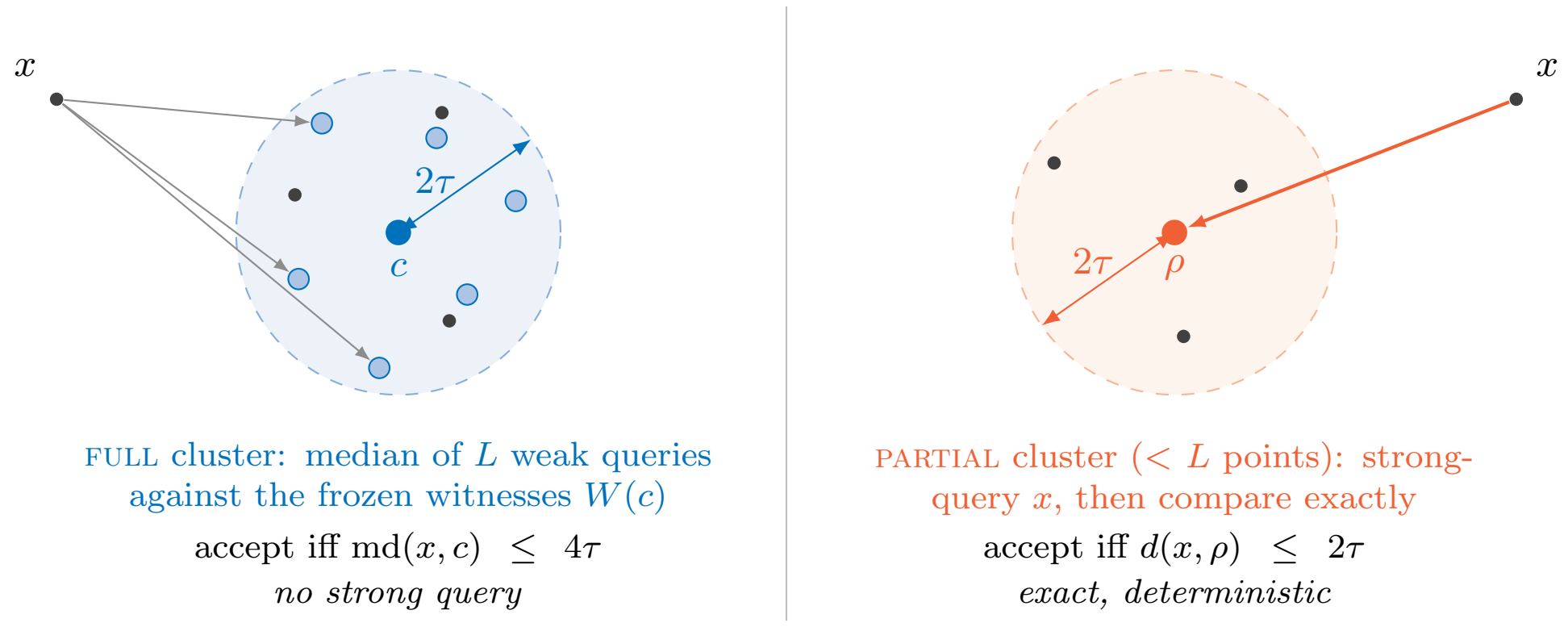
\begin{figure}[t]
\centering
\begin{tikzpicture}[scale=0.95,
   pt/.style={circle,fill=black!75,inner sep=1.1pt},
   rep/.style={circle,fill=RoyalBlue,inner sep=2.1pt},
   wit/.style={circle,draw=RoyalBlue,fill=RoyalBlue!30,inner sep=1.7pt},
   lbl/.style={font=\footnotesize},
   capt/.style={font=\scriptsize,text width=5.3cm,align=center}]

\begin{scope}
  \draw[fill=RoyalBlue!6,draw=RoyalBlue!45,dashed] (0,0) circle (1.4);
  \node[rep] (c) at (0,0) {};
  \node[lbl,text=RoyalBlue] at (0.0,-0.3) {$c$};
  \foreach \a/\r in {15/1.05, 68/0.88, 125/1.15, 205/0.95, 262/1.18, 318/0.80}{
     \node[wit] (w\a) at (\a:\r) {};
  }
  \node[pt] at (160:0.95) {};
  \node[pt] at (300:0.88) {};
  \node[pt] at (70:1.1) {};
  \node[pt] (x) at (-2.95,1.15) {};
  \node[lbl] at (-3.22,1.42) {$x$};
  \foreach \a in {125,205,262}{
     \draw[black!45,thin,-{Latex[length=1.3mm]}] (x) -- (w\a);
  }
  \draw[RoyalBlue,{Latex[length=1.5mm]}-{Latex[length=1.5mm]}] (0,0) -- (35:1.4);
  \node[lbl,text=RoyalBlue,anchor=west] at (0.0,0.45) {$2\tau$};
  \node[capt,text=RoyalBlue] at (0,-2.12) {\safe cluster: median of $L$ weak
        queries against the frozen witnesses $W(c)$};
  \node[capt] at (0,-2.92) {accept iff $\MedDist(x,c)\le4\tau$\\[1pt]
        \emph{no strong query}};
\end{scope}

\draw[black!25,thin] (3.35,-3.35) -- (3.35,1.95);

\begin{scope}[shift={(6.7,0)}]
  \draw[fill=RedOrange!6,draw=RedOrange!45,dashed] (0,0) circle (1.4);
  \node[rep,fill=RedOrange] (r) at (0,0) {};
  \node[lbl,text=RedOrange] at (-0.0,-0.30) {$\rho$};
  \node[pt] at (150:1.2) {};
  \node[pt] at (35:0.7) {};
  \node[pt] at (275:0.9) {};
  \node[pt] (y) at (2.95,1.15) {};
  \node[lbl] at (3.22,1.42) {$x$};
  \draw[RedOrange,thick,-{Latex[length=1.5mm]}] (y) -- (r);
  \draw[RedOrange,{Latex[length=1.5mm]}-{Latex[length=1.5mm]}] (0,0) -- (215:1.4);
  \node[lbl,text=RedOrange] at (-0.7,-0.25) {$2\tau$};
  \node[capt,text=RedOrange] at (0,-2.12) {\uns cluster ($<L$ points):
        strong-query $x$, then compare exactly};
  \node[capt] at (0,-2.92) {accept iff $d(x,\rho)\le2\tau$\\[1pt]
        \emph{exact, deterministic}};
\end{scope}

\end{tikzpicture}
\caption{The two routing rules of $\Test$. \emph{Left:} a \safe cluster carries $L$
frozen witnesses (light blue), each within $2\tau$ of its representative $c$ by
\Cref{lem:witness}; an arriving point $x$ is tested by taking the median of $L$ weak
queries, one per witness. This costs no strong query. \emph{Right:} only if every \safe
test fails is $x$ strong-queried and compared \emph{exactly} against all representatives.}
\label{fig:routing}
\end{figure}

The \Test procedure, \Cref{alg:cover}, is a query-efficient version of Gonzalez's rule.
It processes the points of $U$ one by one and tries to attach each new point $x$ to an existing cluster.
It uses an (expensive) strong query only when the (cheap) weak test against every \safe cluster fails.
Concretely, for each $x$ it first runs the \emph{weak routing} step~(A), which accepts $x$ into a \safe cluster $c$ whenever $\MedDist(x,c)\le4\tau$, using no strong query.
Only if this fails does it run the \emph{exact routing} step~(B): it strong-queries $x$ and compares it with all representatives.
If $x$ is far from all of them, it opens a new cluster (or declares failure).
The two steps are shown in \Cref{fig:routing}.

\begin{algorithm2e}[t]
\caption{$\Test(U,m,\tau,L)$: decide whether $m$ balls of radius $\tau$ cover $U$.}
\label{alg:cover}
\DontPrintSemicolon
$C\gets\emptyset$\;
\ForEach{$x\in U$, in an arbitrary order}{
  \eIf{\textbf{(A)} some \safe $c\in C$ has $\MedDist(x,c)\le 4\tau$}{
    add $x$ to the cluster of $c$\;
  }{
    \textbf{(B)} strong-query $x$ and compute $d(x,\rho)$ for \emph{all} $\rho\in C$\;
    \eIf{some $\rho\in C$ has $d(x,\rho)\le 2\tau$}{
      add $x$ to the cluster of $\rho$\;
    }{
      open a new cluster at $x$, and set $C\gets C\cup\{x\}$\;
      \lIf{$|C|>m$}{\Return \textbf{fail}}
    }
  }
  \lIf{the cluster of $x$ has size exactly $L$}{declare it \safe and freeze its members as $W(\cdot)$}
}
\Return \textbf{pass}, together with the clusters and their representatives\;
\end{algorithm2e}

\subsection{The median test}
\label{sec:median}

Everything the weak oracle contributes is contained in the following lemma.
Note that we need \emph{both} inequalities: the upper bound to route points, the lower bound to conclude in \Cref{lem:sketch} that a routed point really is close to its representative.

\begin{lemma}
\label{lem:median}
Fix $\delta<1/2$, a point $x$, and a set $W$ of $L$ points with $x\notin W$.
With probability at least $1-e^{-2(1/2-\delta)^2L}$ over the weak oracle, strictly more than half of the values $\{\wt d(x,q):q\in W\}$ are uncorrupted, and on that event,
\[
  \min_{q\in W}d(x,q)\ \le\ \median\bigl(\{\wt d(x,q):q\in W\}\bigr)\ \le\
  \max_{q\in W}d(x,q).
\]
\end{lemma}

\begin{proof}
For $q\in W$ let $Z_q=\mathbf 1[\wt d(x,q)=d(x,q)]$.
The pairs $\{x,q\}$, $q\in W$, are distinct, so the $Z_q$ are independent and $Z:=\sum_{q\in W}Z_q\sim\mathrm{Bin}(L,1-\delta)$ with $1-\delta>1/2$.
A Chernoff bound gives $\Pr[Z\le L/2]\le e^{-2(1/2-\delta)^2L}$.

Suppose $Z>L/2$ and sort the $L$ returned values.
More than half of the entries are true distances, so among the $\lceil L/2\rceil$ smallest entries at least one is a true distance, and likewise among the $\lceil L/2\rceil$ largest.
Hence the median entry is at least the smallest true distance and at most the largest one, which is the claim.
\end{proof}

\begin{corollary}
\label{cor:median-center}
On the event of \Cref{lem:median} applied to $W=W(c)$, if every $q\in W(c)$ satisfies $d(c,q)\le2\tau$, then
\[
  d(x,c)-2\tau\ \le\ \MedDist(x,c)\ \le\ d(x,c)+2\tau .
\]
\end{corollary}

\begin{proof}
By the triangle inequality $|d(x,q)-d(x,c)|\le d(c,q)\le2\tau$ for every $q\in W(c)$, so both $\min_q d(x,q)$ and $\max_q d(x,q)$ lie in $[d(x,c)-2\tau,\,d(x,c)+2\tau]$.
Now apply \Cref{lem:median}.
\end{proof}

The hypothesis of \Cref{cor:median-center} is an invariant of \Cref{alg:cover}, and it is worth isolating because it is the only place the order of operations matters.

\begin{lemma}
\label{lem:witness}
In any execution of $\Test(U,m,\tau,L)$, every witness $q\in W(c)$ satisfies $d(c,q)\le2\tau$, and every witness has been strong-queried.
\end{lemma}

\begin{proof}
Step (A) only ever adds points to \safe clusters.
Hence, when a cluster reaches size $L$ and is declared \safe, each of its $L$ members is the representative or joined through step (B).
Step (B) strong-queries a point and admits it only when its \emph{exact} distance to the representative is at most $2\tau$.
The witness set is frozen at that moment, so the property persists.
\end{proof}

We now show that $\Test$ is correct.

\begin{theorem}
\label{thm:cover}
For every $\tau>0$, the following guarantees hold (probability is over weak-oracle corruption):
\begin{enumerate}[label=(\alph*),topsep=2pt]
  \item \emph{(Soundness)} If $\tau < \optC(U,m)/6$, then $\Test(U,m,\tau,L)$ returns {\bf fail} with high probability.
  \item \emph{(Completeness)} If $\tau \geq \optC(U,m)$, then $\Test(U,m,\tau,L)$ always returns {\bf pass}.
\end{enumerate}
\end{theorem}

\begin{proof}
In this proof, we abbreviate $\optC(U,m)$ by $\optC$. Suppose $\tau < \optC/6$. Consider the event in which the algorithm returns {\bf pass}, and let $C$ denote the returned representatives. Since any clustering has radius at least $\optC$, there is some $x \in U$ with $d(x,c) \geq \optC > 6\tau$. It cannot be that $x$ was strong-queried or else the algorithm would have added $x$ to $C$. So $\MedDist(x,c) \leq 4\tau$ for some $c \in C'$ where $C' \subseteq C$ is the set of representatives at the time $x$ was considered.

If the events in \Cref{lem:median} held when $\MedDist(x,c)$ was calculated, then for some $q \in W(c)$  we have $d(x,c) \leq d(x,q) + d(q,c) \leq \MedDist(x,c) + 2\tau \leq 4\tau + 2\tau = 6\tau$. A contradiction. So it must have been that the events in \Cref{lem:median} failed to occur, which happens with probability $n^{-\Omega(1)}$.

In fact, the algorithm performs only polynomially many median estimates, at most $n \cdot m$. So by the union bound, for a large-enough constant in the definition $L = \Theta(\delta^{-2} \log n)$ we have all applications of \Cref{lem:median} are successful. Therefore, if $\tau < \optC/6$ then the algorithm returns {\bf pass} with probability $n^{-\Omega(1)}$ as required.

Now suppose $\tau \ge \optC$ and, by way of contradiction, that the algorithm returns {\bf fail}. In this case, we have $|C| > m$ and $d(x,y) > 2\tau$ for every pair of distinct $x,y \in C$. This is a contradiction since there is some optimal clustering that would have two $x,y \in C$ in the same cluster with center, say, $c^*$ so $d(x,y) \leq d(x,c^*) + d(y,c^*) \leq 2 \optC \leq 2\tau$. So if $\tau \ge \optC$ then the algorithm will always return {\bf pass}.
\end{proof}

Next, we bound the number of strong queries.
Note that the following bound holds for \emph{every} $\tau$, not only for $\tau\ge\optC(U,m)$ which matters, because the search of \Cref{sec:search} runs $\Test$ on failing guesses too.

\begin{lemma}
\label{lem:queries}
A run of $\Test(U,m,\tau,L)$ strong-queries at most $mL+1$ points, w.p. $1-n^{-\Omega(1)}$.
\end{lemma}

\begin{proof}
Let $\mathcal E$ be the event that the conclusion of \Cref{lem:median} holds for every pair $(x,W(c))$ on which step (A) is evaluated.
Only \safe clusters carry a witness set, and at most $m$ clusters ever become \safe.
Indeed, a cluster is frozen only when it reaches size $L$ in the promotion step, which runs \emph{after} the failure check.
So on the iteration that would create an $(m+1)$-st cluster, the procedure returns \textbf{fail} before that cluster can be promoted.
Hence at most $m$ witness sets are ever created.
At most $|U|$ points are processed and each is tested in step~(A) against these $m$ witness sets, so a union bound over the at most $|U|\,m$ median tests gives $\Pr[\overline{\mathcal E}]\ \le\ |U|\,m\,e^{-2(1/2-\delta)^2L}$.
For $L=\Theta((1/2-\delta)^{-2} \log n)$ with a large enough constant this is $n^{-\Omega(1)}$, so $\mathcal E$ holds with high probability. Condition on $\mathcal E$.

A point is strong-queried exactly when it reaches step (B).
We first claim that on $\mathcal E$, any such point $x$ satisfies $d(x,c)>2\tau$ for every \safe representative $c$.
Indeed, if $d(x,c)\le2\tau$ for some \safe $c$, then \Cref{lem:witness} and \Cref{cor:median-center} give $\MedDist(x,c)\le d(x,c)+2\tau\le4\tau$, so step (A) would have accepted $x$ and it would never have reached step (B).

Consequently every point reaching step (B) either opens a new cluster or joins a cluster that is still \uns at that moment.
At most $m+1$ clusters are ever opened; the $(m+1)$-st, if it is opened at all, is the one whose opening triggers the failure, so it contributes exactly one step-(B) point and no joins.
Each of the remaining $\le m$ clusters receives step-(B) points only while \uns, namely the point that opens it and at most $L-1$ further points before it reaches size $L$.
Summing, at most $mL+1$ points reach step~(B).
\end{proof}

\subsection{Searching for the radius, and the sketch}
\label{sec:search}

\begin{lemma}
\label{lem:search}
Assume $|U|>m$.
Binary search over a grid of ratio $1+\eps$, using $O(\log(\eps^{-1}\log n))$ executions of $\Test$, returns a threshold $\tau^\star$ for which $\Test(U,m,\tau^\star,L)$ passes and $\tau^\star \le (1+\eps) \optC(U,m)$.

\end{lemma}

\begin{proof}
Let $G=\{(1+\eps)^t\cdot D/(4\Delta) : t=0,1,\dots,\lceil\log_{1+\eps}(4\Delta^2)\rceil\}$ be a geometric grid of ratio $1+\eps$.
It can be computed from the known $D$ and $\Delta=\poly(n)$ alone, has $O(\eps^{-1}\log n)$ elements, and spans $[d_{\min}/4,\,d_{\max}]$.
Here $d_{\min}$ and $d_{\max}$ are the smallest and largest interpoint distances of $X$.
Its largest element passes \Test, since $\tau\ge d_{\max}\ge\optC(U,m)$ triggers \Cref{thm:cover}(b).
Its smallest element fails \Test, since $\tau<d_{\min}/2$ makes every two points more than $2\tau$ apart.
So every point opens its own cluster and $|C|=|U|>m$ (this is where we use $|U|>m$).

We binary-search for two \emph{adjacent} grid elements $g<g'=(1+\eps)g$ with $g$ failing and $g'$ passing, and set $\tau^\star=g'$.
Such a pair exists and is found in $O(\log|G|)=O(\log(\eps^{-1}\log n))$ tests by the following standard argument, which needs no monotonicity of the pass/fail pattern.
Maintain an interval $[\,\ell,h\,]$ of grid indices with $\Test$ failing at $\ell$ and passing at $h$; the two endpoints of $G$ give such an interval to start.
While $h-\ell>1$, test the midpoint: if it passes, move $h$ to it, otherwise move $\ell$ to it.
Either way the invariant ({\em fail at $\ell$, pass at $h$}) is preserved, so the search halts with $h=\ell+1$, an adjacent fail/pass pair.
Because the grid has ratio $1+\eps$, the grid element below $\tau^\star=g'$ is exactly $g'/(1+\eps)=\tau^\star/(1+\eps)$, so this pair is precisely ``$\Test(U,m,\tau^\star,L)$ passes while $\Test(U,m,\tau^\star/(1+\eps),L)$ fails''.
    
Finally, the run at $\tau^\star/(1+\eps)$ fails.
Since \Cref{thm:cover}(b) guarantees a run at any $\tau\ge\optC(U,m)$ always passes, a fail implies $\tau^\star/(1+\eps)<\optC(U,m)$, i.e.\ $\tau^\star\le (1+\eps)\,\optC(U,m)$, as claimed.
\end{proof}

We can now state the sketch guarantee, which is what the rest of the paper concerns.

\begin{lemma}
\label{lem:sketch}
Let $\Sketch(U,m,L)$ run the search of \Cref{lem:search}, obtaining the threshold $\tau^\star$ together with the passing execution $\Test(U,m,\tau^\star,L)$ that certifies it, and return the clustering that execution produces.
With high probability, it outputs a partition $U=X_{\rho_1}\sqcup\dots\sqcup X_{\rho_{m'}}$ with $m'\le m$ such that
\begin{enumerate}[label=(\roman*),topsep=2pt,itemsep=1pt]
  \item $\tau^\star\le(1+\eps)\,\optC(U,m)$;
  \item every representative $\rho_i$ has been strong-queried, so all distances
        $d(\rho_i,\rho_j)$ are known exactly;
  \item $d(x,\rho_i)\le6\tau^\star$ for every $x\in X_{\rho_i}$;
\end{enumerate}
and it makes $O(mL\log(\eps^{-1}\log n))$ strong queries in total.
\end{lemma}

\begin{proof}
Item (i) is \Cref{lem:search}.
For (ii), every representative is created in step (B), which strong-queries it.
For (iii), a point admitted in step (B) satisfies $d(x,\rho)\le2\tau^\star$ outright, while a point admitted in step (A) against a \safe $c$ has $\MedDist(x,c)\le4\tau^\star$, so on the event of \Cref{lem:median} the lower bound of \Cref{cor:median-center} gives $d(x,c) \le \MedDist(x,c)+2\tau^\star\le 6\tau^\star$.
The query count is \Cref{lem:queries} summed over the $O(\log(\eps^{-1}\log n))$ executions, and the failure probability is a union bound over the same executions.
\end{proof}

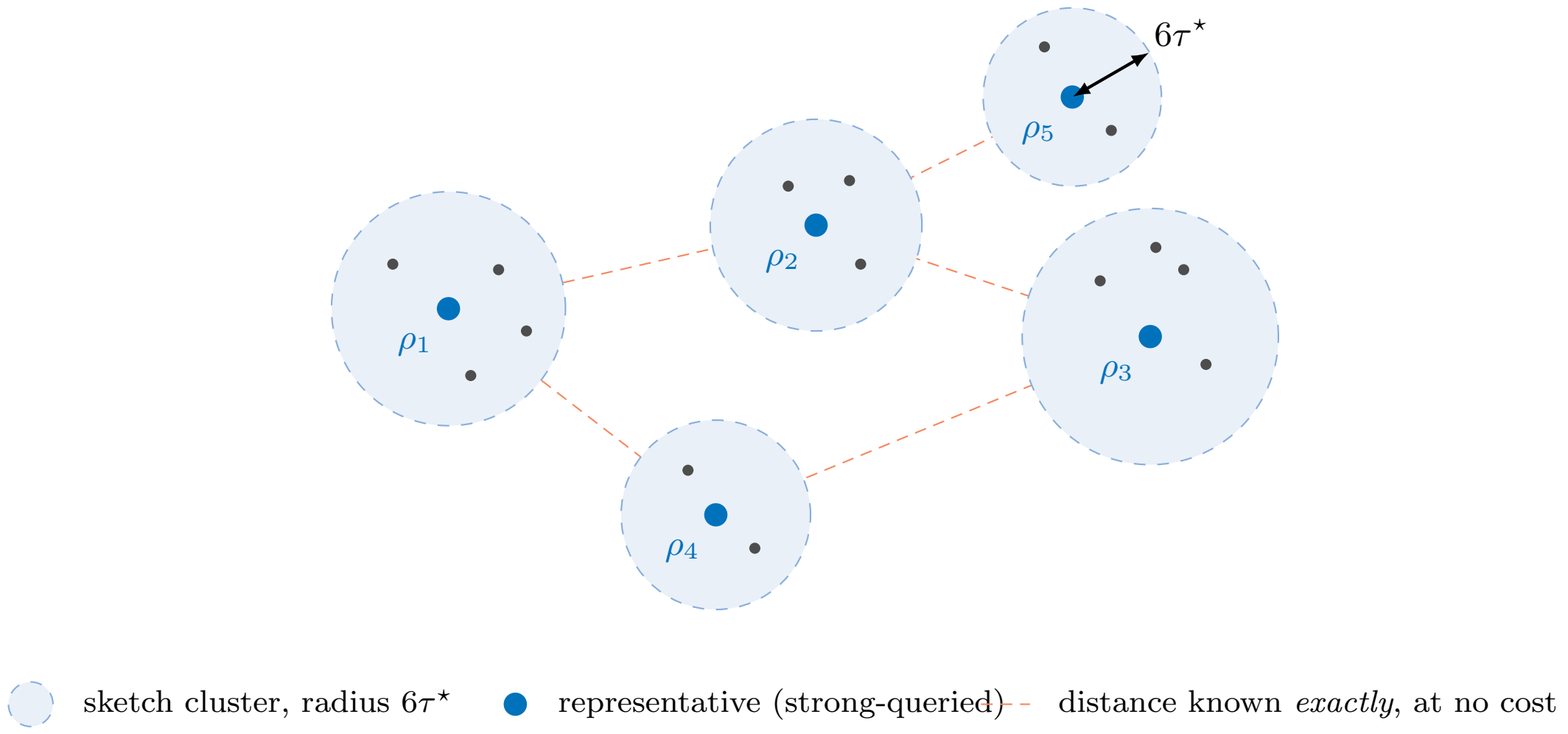
\begin{figure}[t]
\centering
\begin{tikzpicture}[scale=1.0,
   pt/.style={circle,fill=black!70,inner sep=1.0pt},
   rep/.style={circle,fill=RoyalBlue,inner sep=2.1pt},
   lbl/.style={font=\footnotesize},
   leg/.style={font=\scriptsize,anchor=west}]

\begin{scope}[on background layer]
  \draw[RedOrange!75,dash pattern=on 3pt off 2pt] (0,0) -- (3.3,0.75);
  \draw[RedOrange!75,dash pattern=on 3pt off 2pt] (3.3,0.75) -- (6.3,-0.25);
  \draw[RedOrange!75,dash pattern=on 3pt off 2pt] (0,0) -- (2.4,-1.85);
  \draw[RedOrange!75,dash pattern=on 3pt off 2pt] (3.3,0.75) -- (5.6,1.9);
  \draw[RedOrange!75,dash pattern=on 3pt off 2pt] (2.4,-1.85) -- (6.3,-0.25);
\end{scope}

\foreach \x/\y/\R in {0/0/1.05, 3.3/0.75/0.95, 6.3/-0.25/1.15, 2.4/-1.85/0.85, 5.6/1.9/0.8}{
   \draw[fill=RoyalBlue!7,draw=RoyalBlue!45,dashed] (\x,\y) circle (\R);
}
\foreach \x/\y/\name in {0/0/1, 3.3/0.75/2, 6.3/-0.25/3, 2.4/-1.85/4, 5.6/1.9/5}{
   \node[rep] at (\x,\y) {};
}
\node[lbl,text=RoyalBlue] at (-0.30,-0.32) {$\rho_1$};
\node[lbl,text=RoyalBlue] at (3.00,0.43)   {$\rho_2$};
\node[lbl,text=RoyalBlue] at (6.00,-0.57)  {$\rho_3$};
\node[lbl,text=RoyalBlue] at (2.10,-2.17)  {$\rho_4$};
\node[lbl,text=RoyalBlue] at (5.30,1.58)   {$\rho_5$};

\foreach \x/\y in {0.45/0.35, -0.5/0.4, 0.2/-0.6, 0.7/-0.2,
                   3.6/1.15, 3.7/0.4, 3.05/1.1,
                   6.6/0.35, 6.8/-0.5, 5.85/0.25, 6.35/0.55,
                   2.15/-1.45, 2.75/-2.15,
                   5.35/2.35, 5.95/1.6}{
   \node[pt] at (\x,\y) {};
}

\draw[black,thick,{Latex[length=1.6mm]}-{Latex[length=1.6mm]}]
      (5.6,1.9) -- ++(30:0.8);
\node[lbl,text=black,anchor=west] at (6.2,2.48) {$6\tau^\star$};

\begin{scope}[shift={(-3.75,-3.55)}]
  \draw[fill=RoyalBlue!7,draw=RoyalBlue!45,dashed] (0,0) circle (0.20);
  \node[leg] at (0.34,0) {sketch cluster, radius $6\tau^\star$};
  \node[rep] at (4.35,0) {};
  \node[leg] at (4.60,0) {representative (strong-queried)};
  \draw[RedOrange!75,dash pattern=on 3pt off 2pt] (8.53,0) -- (8.97,0);
  \node[leg] at (9.09,0) {distance known \emph{exactly}, at no cost};
\end{scope}

\end{tikzpicture}
\caption{The sketch of \Cref{lem:sketch}. After $O(kL\log\log n)$ strong queries the
algorithm holds $m'\le m$ representatives whose mutual distances it knows exactly,
together with the guarantee that every input point lies within $6\tau^\star=O(\optC)$ of
its own representative.}
\label{fig:sketch}
\end{figure}

\begin{corollary}
\label{cor:kcenter}
Taking $m=k$ and $L=\lceil c\log n\rceil$ for a large enough constant $c=c(\delta)$, $\Sketch(X,k,L)$ returns \whp a $k$-center solution of radius at most $(6+\eps)\,\optC(X,k)$ using $O(k\log n\log(\eps^{-1}\log n))$ strong point queries.
\end{corollary}

\begin{proof}
Immediate from \Cref{lem:sketch}: the representatives form a $k$-center solution of radius $6\tau^\star\le6(1+\eps/6)\,\optC(X,k)=(6+\eps)\,\optC(X,k)$, and the failure probability
is $n^{-\Omega(1)}$ once $c$ is large enough.
\end{proof}

{\section{An $\Omega(k\log(n/k))$ lower bound for row-local algorithms}
\label{sec:rowlocal}

Our algorithm uses $O(k\log n\log\log n)$ strong queries, while the lower bound of~\cite{bateni2024metric} is only $\Omega(k)$.
In this section, we show that the extra $\log n$ factor is needed by every algorithm that assigns an unqueried point using only its weak distances to the queried points.

\subsection{The class of algorithms}
\label{sec:rowlocal-model}

In this section the output is a set $C$ of at most $k$ centers \emph{together with} an assignment $\sigma:X\to C$, and its cost is $\max_{x\in X}d(x,\sigma(x))$.
Let $S$ be the set of strong-queried points.
For $x\notin S$, the \emph{row} of $x$ is the vector $r_S(x):=(\wt d(x,s))_{s\in S}$.

\begin{definition}
\label{def:rowlocal}
An algorithm is \emph{row-local} if it treats any two points $x,y\notin S\cup C$ with $r_S(x)=r_S(y)$ as the same point, that is, $\sigma(x)=\sigma(y)$.
\end{definition}

In other words, if the weak distances of $x$ and $y$ to all points in $S$ are exactly the same, the algorithm assigns $x$ and $y$ to the same center. There is no other restriction: $S$ and $C$ may be chosen using all weak answers.

\begin{proposition}
\label{prop:ours-rowlocal}
Run $\Sketch(U,m,L)$ and then re-assign every $x\in U\setminus S$ to the first \safe representative $c_j$ of the final execution with $\MedDist(x,c_j)\le4\tau^\star$.
The result is row-local and still satisfies \Cref{lem:sketch}(i)--(iii).
\end{proposition}

\begin{proof}
A point $x\notin S$ was accepted in step (A) by some \safe $c$, and $c$ is still \safe at the end with the same witness set, so every $x\notin S$ has at least one such $c_j$.
By \Cref{lem:witness}, all witnesses are in $S$, so the new assignment of $x$ is a deterministic function of $r_S(x)$.
Items (i) and (ii) do not depend on the assignment, and (iii) follows from \Cref{cor:median-center} exactly as in \Cref{lem:sketch}.
\end{proof}

The $k$-center algorithms of~\cite{bateni2024metric} and~\cite{pradhan2026improved} are also row-local.
Both work in rounds.
In each round, they strong-query a random sample of the remaining points.
They use exact distances inside the sample to pick centers, and each center comes with a set of nearby sampled points.
A remaining point $x$ is assigned to a center if the median of its weak distances to these sampled points is small.
The assigned points are then removed.
So the round in which $x$ is assigned, and its center, depend only on $r_S(x)$.
The last step of~\cite{bateni2024metric} only merges centers, so it keeps this property.
If $x$ passes the test for several centers in the same round, we break ties by a fixed order of the centers.
The algorithm of~\cite{pradhan2026improved} uses strong edge queries, and for it we let $S$ be the set of points that appear in a strong query.
As $q$ edge queries involve at most $2q$ points, \Cref{thm:rowlocal} also gives a lower bound of $\Omega(k\log(n/k))$ strong edge queries for row-local algorithms.

We note that this lower bound is for $k$-center with assignment only.
It does not hold if the algorithm only has to output the centers.
On the instance below, strong-querying $O(k\log k)$ random points hits every cluster with constant probability.
Then one queried point per cluster is an optimal set of centers.
The hard part is to assign every point to the correct center.

\subsection{The hard instance}
\label{sec:lb-instance}

Let $k$ be even, $N=n/k$ be an integer, and $\eps\in(0,1)$.
The points are split into $k/2$ pairs, and pair $i$ consists of two \emph{clusters} $P_i^1$ and $P_i^2$ of size $N$, which we call \emph{twins}.
The algorithm knows the union $P_i^1\cup P_i^2$ of every pair, but not how it is split into the two clusters; the split is uniformly random.
Two points in the same cluster are at distance $\eps$, and any other two points are at distance $1$; this is a metric and $\optC(X,k)=\eps$.
So a solution has cost less than $1$ if and only if every point is assigned to a center in its own cluster.

The adversary is as follows.
If the weak answer between two points in the same cluster is corrupted, it reports $1$; if the weak answer between two points in twin clusters is corrupted, it reports $\eps$.
So within two twin clusters, a corrupted weak answer is never the true distance.
Any other two points are reported at distance $1$, their true distance.
See \Cref{fig:lb-instance}.

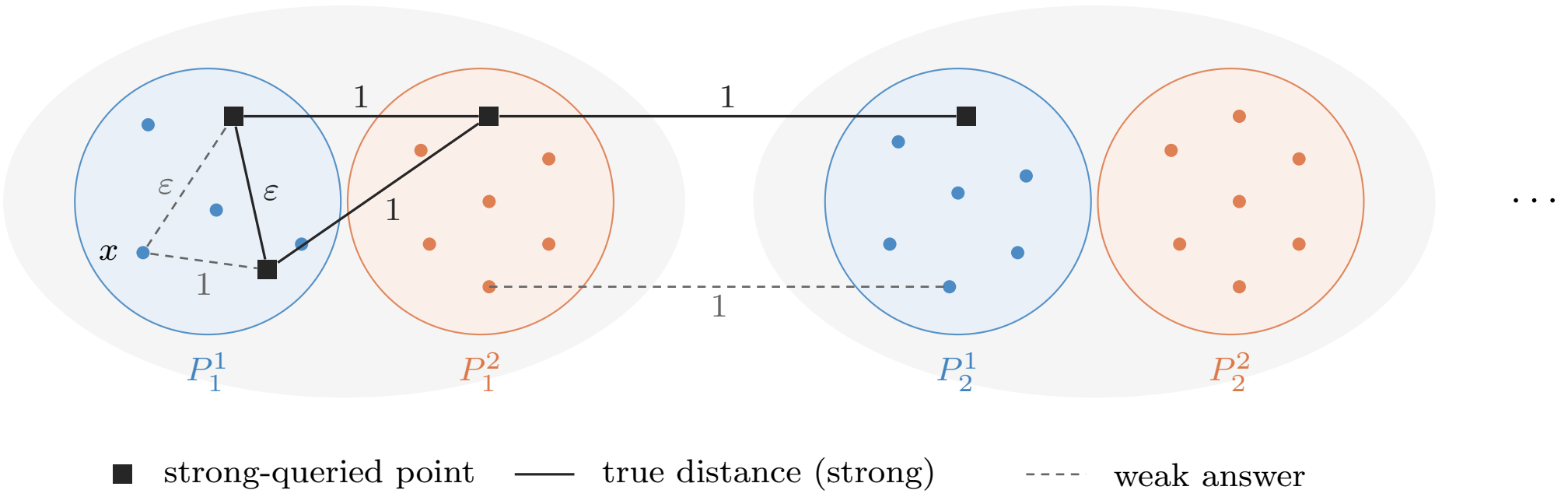
\begin{figure}[t]
\centering
\begin{tikzpicture}[x=1.45cm,y=1.45cm]
  \tikzset{
    ptA/.style={circle,fill=clA!80,inner sep=0pt,minimum size=3.2pt},
    ptB/.style={circle,fill=clB!80,inner sep=0pt,minimum size=3.2pt},
    qr/.style={rectangle,fill=black!85,inner sep=0pt,minimum size=4.6pt},
    weak/.style={black!60,dash pattern=on 2pt off 1.6pt,line width=0.5pt},
    strong/.style={black!85,line width=0.6pt},
    lbl/.style={font=\scriptsize},
    num/.style={font=\scriptsize,inner sep=0.8pt},
    on blue/.style={fill=clA!10},
    on orange/.style={fill=clB!10},
    on grey/.style={fill=black!4},
    on white/.style={fill=white},
  }
  \foreach \xs/\i in {0/1, 4.4/2} {
    \begin{scope}[shift={(\xs,0)}]
      \fill[black!4] (1.6,0) ellipse (2.0 and 1.15);
      \draw[clA!75,fill=clA!10] (0.8,0) circle (0.78);
      \draw[clB!75,fill=clB!10] (2.4,0) circle (0.78);
      \foreach \p in {(2.05,0.3),(2.45,0.5),(2.8,0.25),(2.1,-0.25),(2.45,-0.5),(2.8,-0.25),(2.45,0.0)}
        \node[ptB] at \p {};
      \node[lbl,text=clA!85] at (0.8,-1.0) {$P_{\i}^1$};
      \node[lbl,text=clB!85] at (2.4,-1.0) {$P_{\i}^2$};
    \end{scope}
  }
  \foreach \p in {(0.45,0.45),(0.85,-0.05),(1.35,-0.25)} \node[ptA] at \p {};
  \foreach \p in {(4.85,0.35),(5.25,0.5),(5.6,0.15),(4.8,-0.25),(5.15,-0.5),(5.55,-0.3),(5.2,0.05)} \node[ptA] at \p {};
  \node[qr] (q1) at (0.95,0.5) {};
  \node[qr] (q2) at (1.15,-0.4) {};
  \node[qr] (q3) at (2.45,0.5) {};
  \node[qr] (q4) at (5.25,0.5) {};
  \draw[strong] (q1) -- node[num,on blue,right=2pt] {$\eps$} (q2);
  \draw[strong] (q2) -- node[num,on orange,below right=0.5pt] {$1$} (q3);
  \draw[strong] (q1) -- node[num,on grey,above=1pt] {$1$} (q3);
  \draw[strong] (q3) -- node[num,on white,above=1pt] {$1$} (q4);
  \node[ptA] (x) at (0.42,-0.3) {};
  \node[lbl,anchor=east] at (0.37,-0.3) {$x$};
  \draw[weak] (x) -- node[num,on blue,left=2pt] {$\eps$} (q1);
  \draw[weak] (x) -- node[num,on blue,below=2pt] {$1$} (q2);
  \draw[weak] (2.45,-0.5) -- node[num,on white,below=1pt] {$1$} (5.15,-0.5);
  \node[font=\small] at (8.6,0) {$\cdots$};
  \begin{scope}[shift={(0.3,-1.6)}]
    \node[qr] at (0,0) {};
    \node[lbl,anchor=west] at (0.15,0) {strong-queried point};
    \draw[strong] (2.3,0) -- (2.65,0);
    \node[lbl,anchor=west] at (2.72,0) {true distance (strong)};
    \draw[weak] (5.3,0) -- (5.65,0);
    \node[lbl,anchor=west] at (5.72,0) {weak answer};
  \end{scope}
\end{tikzpicture}
\caption{The hard instance: $k/2$ pairs, each with two twin clusters $P_i^1,P_i^2$ of size $N$. The algorithm knows each grey region $P_i^1\cup P_i^2$, but not how it is split into the two clusters. True distances are $\eps$ inside a cluster and $1$ otherwise. Here the weak answer between $x$ and the second queried point of its cluster is corrupted ($1$ instead of $\eps$).}
\label{fig:lb-instance}
\end{figure}

\smallskip
We say that $x$ and $y$ \emph{look close} if $\wt d(x,y)=\eps$.
\begin{observation}
\label{obs:sbm}
Two points in the same cluster look close with probability $1-\delta$, two points in twin clusters look close with probability $\delta$, all independently, and any other two points never look close.
\end{observation}

\subsection{Points with the same row}
\label{sec:lb-rows}

The whole lower bound comes from the following simple fact.

\begin{lemma}
\label{lem:collision}
Fix a pair $i$, and let $x\in P_i^1$ and $y\in P_i^2$ be two unqueried points.
If all weak answers between $x$ and the queried points of $P_i^1\cup P_i^2$ are corrupted, and all weak answers between $y$ and these points are correct, then $r_S(x)=r_S(y)$.
\end{lemma}

\begin{proof}
Let $s\in S$ be a queried point. We show that $x$ and $y$ either both look close to $s$ or both look far from it.
If $s\in P_i^2$, then $y$ looks close to $s$, since $d(y,s)=\eps$ and the answer for $\{y,s\}$ is correct; so does $x$, since the answer for $\{x,s\}$ is corrupted and the adversary reports every corrupted partnered pair as close.
If $s\in P_i^1$, then $y$ looks far from $s$, since $d(y,s)=1$ and the answer for $\{y,s\}$ is correct; so does $x$, since the answer for $\{x,s\}$ is corrupted and the adversary reports every corrupted same-cluster pair as far.
If $s\notin P_i^1\cup P_i^2$, then $s$ looks far to both $x$ and $y$, since the weak answer between points of different pairs is never $\eps$.
Hence $x$ and $y$ look close to exactly the queried points of $P_i^2$, and $r_S(x)=r_S(y)$.
\end{proof}
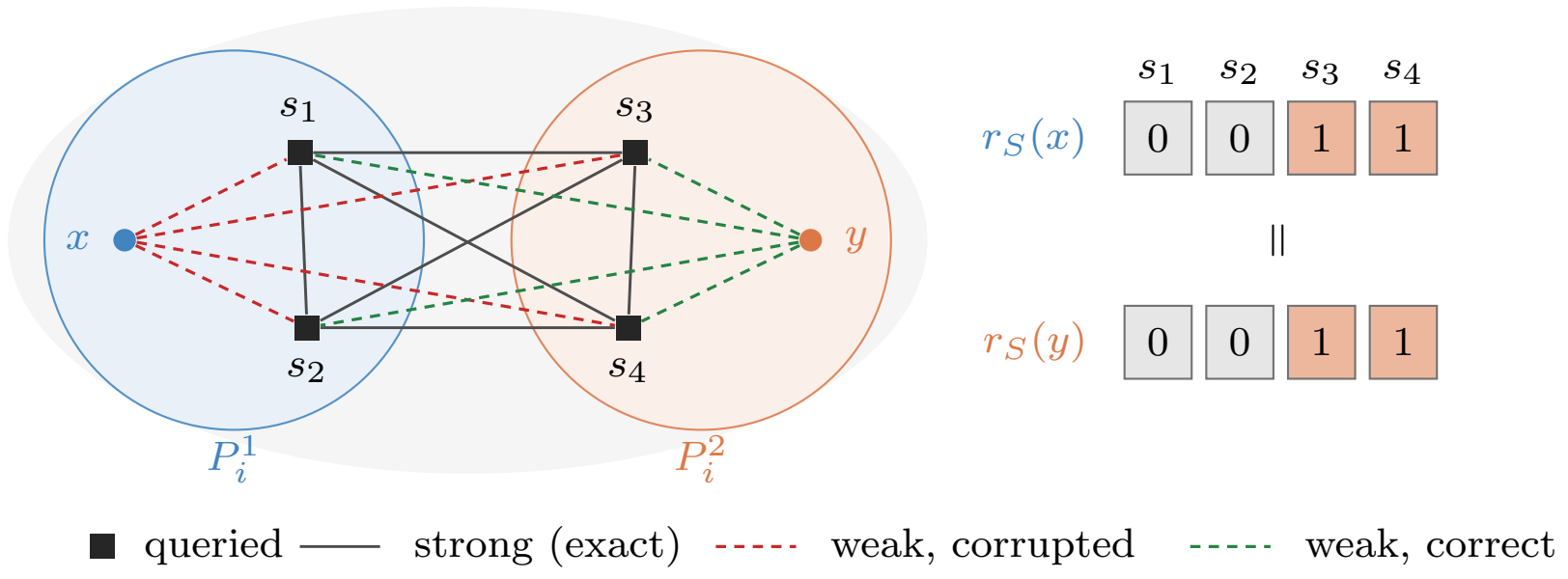
\begin{figure}[t]
\centering
\begin{tikzpicture}[x=1cm,y=1cm]
  \tikzset{
    pt/.style={circle,inner sep=0pt,minimum size=4.4pt},
    qr/.style={rectangle,fill=black!85,inner sep=0pt,minimum size=4.8pt},
    strong/.style={black!70,line width=0.55pt},
    lbl/.style={font=\scriptsize},
  }
  \fill[black!4] (1.6,0) ellipse (3.15 and 1.6);
  \draw[clA!75,fill=clA!10] (0,0) circle (1.3);
  \draw[clB!75,fill=clB!10] (3.2,0) circle (1.3);
  \node[lbl,text=clA!85] at (0,-1.5) {$P_i^1$};
  \node[lbl,text=clB!85] at (3.2,-1.5) {$P_i^2$};
  \node[qr] (s1) at (0.45,0.6) {};   \node[lbl,anchor=south] at (0.45,0.68) {$s_1$};
  \node[qr] (s2) at (0.5,-0.6) {};   \node[lbl,anchor=north] at (0.5,-0.68) {$s_2$};
  \node[qr] (s3) at (2.75,0.6) {};   \node[lbl,anchor=south] at (2.75,0.68) {$s_3$};
  \node[qr] (s4) at (2.7,-0.6) {};   \node[lbl,anchor=north] at (2.7,-0.68) {$s_4$};
  \draw[strong] (s1) -- (s2) -- (s4) -- (s3) -- (s1);
  \draw[strong] (s1) -- (s4);
  \draw[strong] (s2) -- (s3);
  \node[pt,fill=clA!85] (x) at (-0.75,0) {};
  \node[lbl,text=clA!85,anchor=east] at (-0.85,0) {$x$};
  \node[pt,fill=clB!85] (y) at (3.95,0) {};
  \node[lbl,text=clB!85,anchor=west] at (4.05,0) {$y$};
  \foreach \s in {s1,s2,s3,s4}
    \draw[badred,dash pattern=on 2pt off 1.6pt,line width=0.6pt] (x) -- (\s);
  \foreach \s in {s1,s2,s3,s4}
    \draw[okgreen,dash pattern=on 2pt off 1.6pt,line width=0.6pt] (y) -- (\s);
  \begin{scope}[shift={(6.1,0)}]
    \foreach \j/\v/\c in {0/0/black!10, 1/0/black!10, 2/1/clB!45, 3/1/clB!45} {
      \draw[draw=black!55,fill=\c] (\j*0.56,0.45) rectangle (\j*0.56+0.46,0.95);
      \node[lbl] at (\j*0.56+0.23,0.7) {$\v$};
      \draw[draw=black!55,fill=\c] (\j*0.56,-0.95) rectangle (\j*0.56+0.46,-0.45);
      \node[lbl] at (\j*0.56+0.23,-0.7) {$\v$};
      \node[lbl] at (\j*0.56+0.23,1.15) {$s_{\the\numexpr\j+1\relax}$};
    }
    \node[lbl,text=clA!85,anchor=east] at (-0.12,0.7) {$r_S(x)$};
    \node[lbl,text=clB!85,anchor=east] at (-0.12,-0.7) {$r_S(y)$};
    \node[font=\footnotesize] at (1.07,0) {\rotatebox{90}{$=$}};
  \end{scope}
  \begin{scope}[shift={(-0.9,-2.1)}]
    \node[qr] at (0,0) {};
    \node[lbl,anchor=west] at (0.15,0) {queried};
    \draw[strong] (1.35,0) -- (1.9,0);
    \node[lbl,anchor=west] at (2.0,0) {strong (exact)};
    \draw[badred,dash pattern=on 2pt off 1.6pt,line width=0.6pt] (4.2,0) -- (4.75,0);
    \node[lbl,anchor=west] at (4.85,0) {weak, corrupted};
    \draw[okgreen,dash pattern=on 2pt off 1.6pt,line width=0.6pt] (7.45,0) -- (8.0,0);
    \node[lbl,anchor=west] at (8.1,0) {weak, correct};
  \end{scope}
\end{tikzpicture}
\caption{\Cref{lem:collision}: all weak answers of $x\in P_i^1$ to the queried points of its pair are corrupted, and all those of $y\in P_i^2$ are correct, so $x$ and $y$ have the same row ($1$ = looks close).}
\label{fig:lb-collision}
\end{figure}

So a row-local algorithm must treat such $x$ and $y$ in the same way, although they are in different clusters.

\begin{lemma}
\label{lem:confusion}
If $x$ and $y$ are in different clusters, $x,y\notin S\cup C$ and $r_S(x)=r_S(y)$, then a row-local algorithm outputs a solution of cost $1$.
\end{lemma}

\begin{proof}
By \Cref{def:rowlocal}, $\sigma(x)=\sigma(y)$.
This center is in at most one of the two clusters, so $x$ or $y$ is at distance $1$ from it.
\end{proof}

It remains to show that, with few queries, some pair contains such $x$ and $y$ that are not centers. Set $\ell:=\lfloor\tfrac12\log_{1/\delta}N\rfloor$.

\begin{lemma}
\label{lem:lb-counting}
Let $\Good$ be the event that for every pair $i$ and every set $T$ of at most $\ell$ points of $P_i^1\cup P_i^2$:
\begin{enumerate}[label=(\roman*),topsep=2pt,itemsep=1pt]
  \item at least $\frac14\sqrt N$ points of $P_i^1\setminus T$ have all weak answers to $T$ corrupted, and
  \item at least $\frac14\sqrt N$ points of $P_i^2\setminus T$ have all weak answers to $T$ correct.
\end{enumerate}
Then $\Pr[\Good]\ge 1-k(2N)^\ell e^{-\sqrt N/16}$.
\end{lemma}

\begin{proof}
The probability is over which weak answers are corrupted, and that is independent of how each pair is split into $P_i^1$ and $P_i^2$.
So we fix the split and prove the bound for every split.
Recall that $\ell\le\frac12\log_{1/\delta}N$.

If a point has all its weak answers to $T$ corrupted, it also has all its weak answers to every $T'\subseteq T$ corrupted, and $P_i^1\setminus T\subseteq P_i^1\setminus T'$; the same holds for correct answers and $P_i^2$.
So if (i) and (ii) hold for $T$, they hold for every $T'\subseteq T$. So it suffices to prove (i) and (ii) for sets $T$ with $|T|=\ell$.

Fix a pair $i$ and a set $T\subseteq P_i^1\cup P_i^2$ with $|T|=\ell$, and let $Z$ be the number of points of $P_i^1\setminus T$ whose weak answers to $T$ are all corrupted.
Take $x\in P_i^1\setminus T$.
Since $x\notin T$, the $\ell$ pairs $\{x,t\}$ with $t\in T$ are distinct.
Each of them is corrupted independently with probability $\delta$, so
\[
  \Pr[\text{all weak answers between $x$ and $T$ are corrupted}]
  \ =\ \delta^{\ell}
  \ \ge\ \delta^{\frac12\log_{1/\delta}N}
  \ =\ N^{-1/2}.
\]
For distinct $x,x'\in P_i^1\setminus T$ these sets of pairs are disjoint, so these events are independent, and $Z$ is a sum of $|P_i^1\setminus T|$ independent indicator variables.
Since $\delta<1/2$, we have $\ell\le\frac12\log_2N\le N/2$, so $|P_i^1\setminus T|\ge N-\ell\ge N/2$ and $\mu:=\mathbb E[Z]\ge\frac N2\cdot N^{-1/2}=\frac12\sqrt N$.
By the multiplicative Chernoff bound $\Pr[Z\le(1-\eta)\mu]\le e^{-\eta^2\mu/2}$ with $\eta=\frac12$, and since $\frac14\sqrt N\le\frac\mu2$,
\[
  \Pr\bigl[Z<\tfrac14\sqrt N\bigr]\ \le\ \Pr\bigl[Z\le\tfrac\mu2\bigr]\ \le\ e^{-\mu/8}\ \le\ e^{-\sqrt N/16}.
\]

For (ii), the same argument applies to the points of $P_i^2\setminus T$ whose weak answers to $T$ are all correct: each has this property with probability $(1-\delta)^\ell\ge\delta^\ell\ge N^{-1/2}$, independently of the others for the same reason.

Finally, $\Good$ fails only if (i) or (ii) fails for some pair $i$ and some set $T$ of size $\ell$.
There are $k/2$ pairs and $\binom{2N}{\ell}\le(2N)^\ell$ such sets $T$ for each pair, and for each of them, (i) and (ii) each fail with probability at most $e^{-\sqrt N/16}$.
By the union bound,
\begin{align*}
  1-\Pr[\Good]
  \le \sum_{i=1}^{k/2}\ \sum_{\substack{T\subseteq P_i^1\cup P_i^2\\|T|=\ell}}
         \Bigl(\Pr[\text{(i) fails for $T$}]+\Pr[\text{(ii) fails for $T$}]\Bigr)
  &\le \frac k2\cdot(2N)^\ell\cdot 2e^{-\sqrt N/16} \\
  &= k(2N)^\ell e^{-\sqrt N/16}. 
\end{align*}

\end{proof}

The lemma holds for every set $T$ at once, so we can apply it to the queried points of a pair even though $S$ is chosen adaptively.

\subsection{The lower bound}
\label{sec:lb-main}

\begin{theorem}
\label{thm:rowlocal}
Let $n$ be large enough and $N\ge\log^5n$.
On the instance of \Cref{sec:lb-instance}, every row-local algorithm with at most $k\ell/4$ strong queries has cost $1$ with probability at least $3/4$, while the optimal $k$-center clustering has cost $\eps$.
\end{theorem}

\begin{proof}
For $n$ large enough we have $\ell\ge1$, $\sqrt N>16$ and $\Pr[\Good]\ge3/4$ by \Cref{lem:lb-counting}, because $\ell=O(\log n)$ while $\sqrt N\ge\log^{5/2}n$.
Fix the instance, the corrupted weak answers and the algorithm's choice of $S$ and $C$, and assume $\Good$ holds.

Call pair $i$ \emph{light} if $P_i^1\cup P_i^2$ has at most $\ell$ queried points.
At most $|S|/(\ell+1)<k/4$ pairs are not light, so more than $k/4$ of the $k/2$ pairs are light.

As $|C|\le k$, at most $4k/\sqrt N<k/4$ pairs contain $\frac14\sqrt N$ or more centers, so some light pair $i$ contains fewer than $\frac14\sqrt N$ centers.
Let $T$ be the queried points of $P_i^1\cup P_i^2$, so $|T|\le\ell$.
By (i) of $\Good$, at least $\frac14\sqrt N$ points of $P_i^1\setminus T$ have all weak answers to $T$ corrupted; as fewer than $\frac14\sqrt N$ points of the pair are centers, one of them, $x$, is not a center.
In the same way, by (ii) there is $y\in P_i^2\setminus T$ that is not a center and has all weak answers to $T$ correct.
Both are unqueried, so by \Cref{lem:collision} they have the same row, and by \Cref{lem:confusion} the solution has cost $1$.

Therefore the solution has cost $1$ whenever $\Good$ holds, which is with probability at least $3/4$.
\end{proof}

\begin{corollary}
\label{cor:rowlocal}
Let $\alpha\ge1$ and $c>0$ be constants, and let $k\le n^{1-c}$ be even and divide $n$.
Every row-local algorithm that outputs an $\alpha$-approximate assignment with probability more than $1/4$ makes $\Omega(k\log n)$ strong queries.
\end{corollary}

\begin{proof}
Take $\eps=1/(2\alpha)$, so an $\alpha$-approximate solution has cost less than $1$.
Since $N\ge n^c$, the condition $N\ge\log^5n$ holds for $n$ large enough, and $\ell=\Theta(\log N)=\Theta(\log n)$.
Now apply \Cref{thm:rowlocal}.
\end{proof}

\subsection{The restriction is needed}
\label{sec:lb-barrier}

The lower bound does not hold for all algorithms.
A row-local algorithm assigns a point using only its weak answers to the queried points.
An algorithm that also compares unqueried points with each other solves our instance with no strong queries.

For two points $u,v$ in the same pair $i$, let $\mathrm{cn}(u,v)$ be the number of points of $P_i^1\cup P_i^2$ other than $u,v$ that look close to both $u$ and $v$.
By \Cref{obs:sbm},
\[
  \mathbb E[\mathrm{cn}(u,v)]\ =\
  \begin{cases}
    (N-2)(1-\delta)^2+N\delta^2, & \text{if $u,v$ are in the same cluster,}\\
    (2N-2)\,\delta(1-\delta), & \text{if $u,v$ are in twin clusters.}
  \end{cases}
\]
As $\delta<1/2$, the first value is larger, and the difference between the two is at least $\frac12N(1-2\delta)^2$ once $N\ge4(1-\delta)/(1-2\delta)$.
Let $\theta$ be the midpoint of the two values.

\begin{proposition}
\label{prop:barrier}
Let $N\ge C\log n$ for a large enough constant $C$.
Put two points $u,v$ of the same pair in the same cluster if $\mathrm{cn}(u,v)>\theta$.
With probability at least $1-1/n$, this finds all clusters correctly, using no strong queries.
\end{proposition}

\begin{proof}[Proof sketch]
For fixed $u,v$, $\mathrm{cn}(u,v)$ is a sum of at most $2N$ independent indicators, so by Hoeffding's inequality it is on the correct side of $\theta$ with probability at least $1-2e^{-N(1-2\delta)^4/16}$.
For $N\ge C\log n$ this is at least $1-n^{-3}$, and a union bound over all choices of $u,v$ completes the proof.
\end{proof}

So an unconditional $\omega(k)$ lower bound needs a different hard instance, one in which comparing unqueried points with each other does not reveal the clusters.

}

\section{Application: an FPT constant-factor algorithm for MSR}
\label{sec:fptmsr}

This section applies the main result to MSR and proves \Cref{thm:intro-msr}.
Everything below treats the sketch as a black box.
We only use that there are at most $k$ representatives, that their mutual distances are known exactly, and that every point lies within $6\tau^\star$ of its own representative.
The weak oracle does not appear in this section at all.

\subsection{An $\Omega(k)$ lower bound on strong queries for MSR}
\label{sec:lb}

We first record why $O(k)$ is the right target.
The following adapts the $\Omega(k)$ bound for other $k$-clustering problems of~\cite{bateni2024metric} to MSR; the proof is included for completeness.

\begin{theorem}
\label{thm:lb}
Fix an approximation factor $\alpha\ge1$ and a corruption probability $\delta\in(0,1/2)$.
There are $n$-point MSR instances with $k=n-1$ on which every randomized algorithm making $o(n)$ strong point queries fails to output an $\alpha$-approximation with probability at least $\delta-o(1)$.
\end{theorem}

\begin{proof}
By Yao's principle, it suffices to exhibit a distribution over instances on which every \emph{deterministic} algorithm making $o(n)$ strong queries fails with the stated probability.
Fix a constant $M>\alpha$, choose a pair $\{u,v\}\subseteq X$ uniformly at random, and set
\[
  d(u,v)=1,\qquad d(x,y)=M\ \text{ for every other pair }\{x,y\}.
\]
Every triangle has side lengths $(M,M,M)$ or $(1,M,M)$, so $d$ is a metric for $M\ge1$.
Let the adversary answer every corrupted weak query with the value $M$.

Take $k=n-1$.
Clustering $\{u,v\}$ together and leaving the other $n-2$ points as singletons costs $1$, and any solution with at most $n-1$ clusters must place two points in a common cluster, so $\optR(X,n-1)=1$.
Moreover any solution that puts two points other than $u,v$ in a common cluster has a cluster of radius $M>\alpha=\alpha\cdot\optR(X,n-1)$.
Hence an $\alpha$-approximation must identify the hidden pair exactly.

Now consider the information available.
With probability $\delta$, the weak query for the pair $\{u,v\}$ is corrupted.
In that case every weak answer in the whole instance equals $M$, so the weak oracle gives no information.
So the posterior on $\{u,v\}$ stays uniform over all pairs not yet excluded by strong queries.
A strong query reveals a pair's distance only when \emph{both} endpoints have been queried, so after $q=o(n)$ strong queries at most $\binom q2=o(n^2)$ pairs are excluded, and the algorithm's output is correct with probability $O(q^2/n^2)=o(1)$.
Failure probability is therefore at least $\delta\cdot(1-o(1))=\delta-o(1) =\Omega(1)$.
\end{proof}

\subsection{The algorithm}
\label{sec:fpt-alg}
The algorithm is a branching recursion that removes a single optimal ball at each level.
Its only primitive is the sketch of \Cref{lem:sketch}; after the sketch is built, the weak oracle is not consulted again, and every quantity the algorithm inspects is an exactly known distance between two representatives.

Let $U$ be the current residual with budget $m$, and suppose $|U|>m$ (otherwise output $|U|$ singleton balls of radius $0$ and stop).
Run $\Sketch(U,m,L)$ with $\eps=1$, so that \Cref{lem:sketch}(i) reads $\tau^\star\le2\,\optC(U,m)$, to obtain $\tau:=\tau^\star$, representatives $\rho_1,\dots,\rho_{m'}$ and clusters $X_{\rho_1},\dots,X_{\rho_{m'}}$.
In every execution of $\Test$ inside $\Peel$, a point that was strong-queried in an earlier execution is never tested by the weak oracle.
It is compared directly with the representatives using exact distances, as in step~(B) of \Cref{alg:cover}.
It is also never used as a witness again.
For a representative $\rho$ and a radius guess $R$ define
\begin{equation}
\label{eq:Y}
  Y(\rho,R)\ :=\ \bigcup\bigl\{\,X_{\rho'}\ :\ d(\rho,\rho')\le R+12\tau\,\bigr\}.
\end{equation}
Every distance appearing in \eqref{eq:Y} is between two representatives, hence known exactly and for free by \Cref{lem:sketch}(ii); computing $Y(\rho,R)$ costs no queries at all.

\begin{algorithm2e}[t]
\caption{$\Peel(U,m)$: return a set of at most $m$ balls covering $U$, or \textbf{infeasible}.}
\label{alg:peel}
\DontPrintSemicolon
\lIf{$|U|\le m$}{\Return $\{\Ball(x,0):x\in U\}$}
\lIf{$m=0$}{\Return \textbf{infeasible}}
$(\tau,\{\rho_i\},\{X_{\rho_i}\})\gets\Sketch(U,m,L)$\;
$\mathcal G\gets\bigl\{(\tau/2)\cdot2^j\ :\ j=0,1,\dots,\lceil\log_2(48m)\rceil\bigr\}$ \tcp*{$O(\log k)$ radius guesses}
\ForEach{representative $\rho$ and guess $R\in\mathcal G$}{
  $Y\gets Y(\rho,R)$ as in \eqref{eq:Y}\;
  $\mathcal S\gets\Peel(U\setminus Y,\;m-1)$\;
  \lIf{$\mathcal S\ne\textbf{infeasible}$}{record the candidate $\{\Ball(\rho,\,R+18\tau)\}\cup\mathcal S$}
}
\Return the recorded candidate of minimum total cost, or \textbf{infeasible} if there is none\;
\end{algorithm2e}

\begin{figure}[t]
\centering
\begin{tikzpicture}[scale=0.98,
   rep/.style={circle,fill=RoyalBlue,inner sep=1.7pt},
   ctr/.style={circle,fill=RedOrange,inner sep=2.0pt},
   lbl/.style={font=\footnotesize},
   leg/.style={font=\scriptsize,anchor=west}]

\draw[fill=RoyalBlue!7,draw=RoyalBlue,thick] (0,0) circle (3.25);

\draw[fill=RedOrange!10,draw=RedOrange,thick,dash pattern=on 4pt off 2.5pt]
      (0.45,0.20) circle (1.95);

\foreach \x/\y/\R in {1.60/0.95/0.52, -1.30/0.80/0.50, 0.35/-1.30/0.50,
                      1.80/-0.70/0.46, -1.20/-1.05/0.48, -0.40/1.80/0.46,
                      2.35/0.20/0.44, -2.20/0.10/0.44}{
   \draw[fill=RoyalBlue!25,draw=RoyalBlue!70,thin] (\x,\y) circle (\R);
   \node[rep] at (\x,\y) {};
}

\draw[fill=RoyalBlue!25,draw=RoyalBlue!70,thin] (0,0) circle (0.72);
\node[rep] at (0,0) {};
\node[lbl,text=RoyalBlue] at (-0.30,-0.34) {$\rho$};
\node[ctr] at (0.42,0.30) {};
\node[lbl,text=RedOrange] at (0.72,0.56) {$o^\star$};

\foreach \x/\y/\R in {-3.75/-2.15/0.44, 3.60/-2.45/0.44, 3.05/2.55/0.42}{
   \draw[fill=black!6,draw=black!40,thin,dashed] (\x,\y) circle (\R);
   \node[rep,fill=black!45] at (\x,\y) {};
}

\draw[RoyalBlue,{Latex[length=1.6mm]}-{Latex[length=1.6mm]}] (0,0) -- (300:3.25);
\node[lbl,text=RoyalBlue,fill=RoyalBlue!7,inner sep=1.5pt] at (1.55,-2.30)
      {$R{+}18\tau$};

\begin{scope}[shift={(-4.9,-4.55)}]
  \draw[fill=RoyalBlue!7,draw=RoyalBlue,thick] (0,0) circle (0.20);
  \node[leg] at (0.34,0) {ball added: $\Ball(\rho,\,R{+}18\tau)$};
  \draw[fill=RedOrange!10,draw=RedOrange,thick,dash pattern=on 2.5pt off 1.8pt]
        (5.0,0) circle (0.20);
  \node[leg] at (5.34,0) {largest optimal ball: $\Ball(o^\star,r_{\max})$};

  \draw[fill=RoyalBlue!25,draw=RoyalBlue!70,thin] (0,-0.72) circle (0.20);
  \node[leg] at (0.34,-0.72) {clusters removed: $Y(\rho,R)$};
  \draw[fill=black!6,draw=black!40,thin,dashed] (5.0,-0.72) circle (0.20);
  \node[leg] at (5.34,-0.72) {clusters kept: passed to the next level};
\end{scope}

\end{tikzpicture}
\caption{One level of $\Peel$. The removal set $Y(\rho,R)$ is a union of \emph{whole}
sketch clusters, selected using only the exactly-known distances between representatives. By \Cref{lem:sandwich} it \emph{contains} the largest
optimal ball, so deleting it drops the residual optimum by $r_{\max}$, and it is
\emph{contained} in the single ball of radius $R+18\tau=O(r_{\max})$ that we add to pay
for it. Note that $\rho$ need not be near the center of the optimal ball; it only needs to represent the cluster of a point covered by the ball ($o^\star$ itself need not even lie in the residual set.)}
\label{fig:peel}
\end{figure}
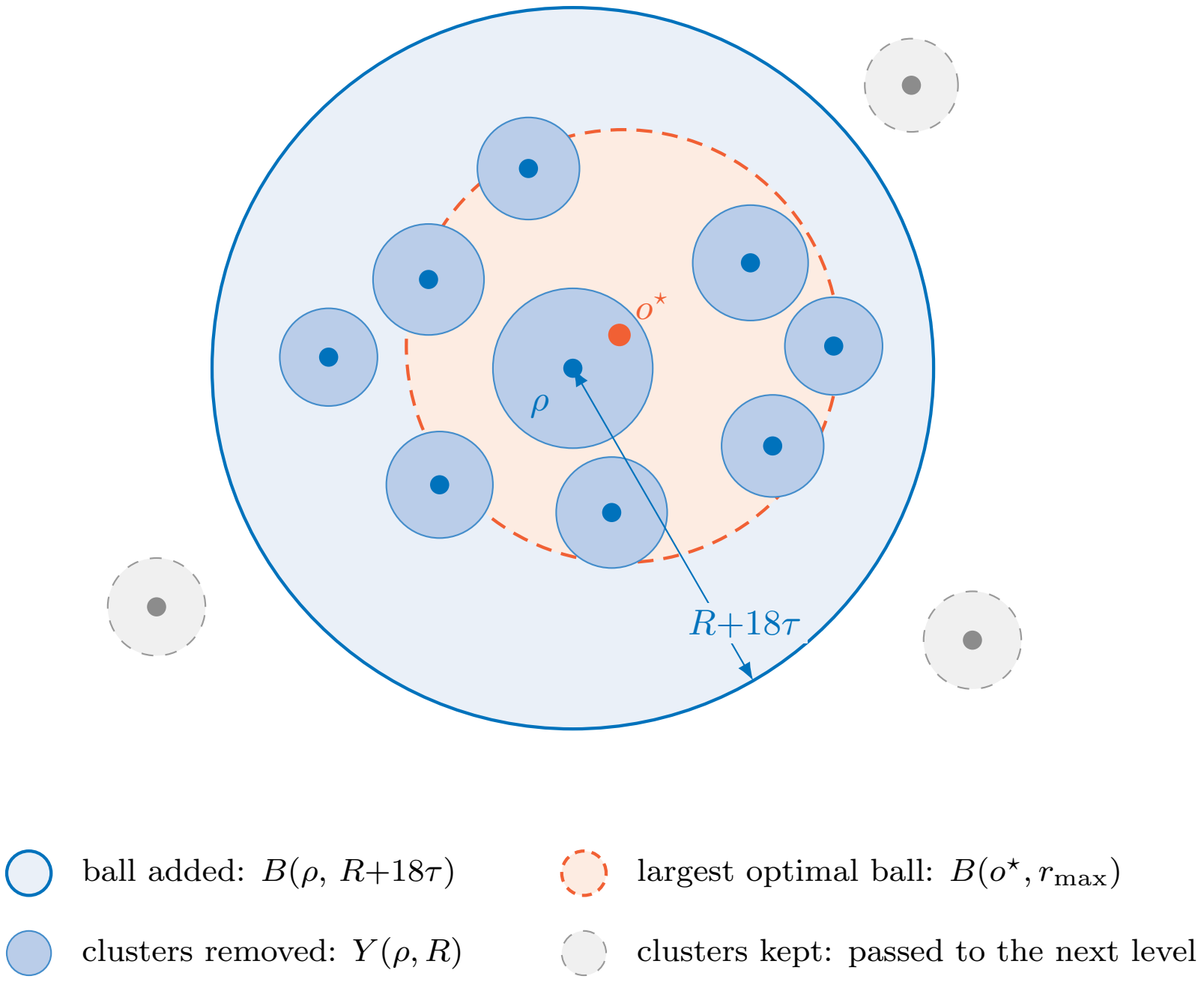

The entire analysis rests on the following two-line consequence of the triangle inequality, illustrated in \Cref{fig:peel}.

\begin{lemma}
\label{lem:sandwich}
Suppose the sketch guarantee of \Cref{lem:sketch}(iii) holds.
Let $o\in X$ and $r\ge0$ be such that $\Ball(o,r)\cap U\ne\emptyset$, let $z$ be any point of $\Ball(o,r)\cap U$, let $\rho$ be the representative of the cluster containing $z$, and let $R\ge 2r$. Then $ \Ball(o,r)\cap U\ \subseteq\ Y(\rho,R)\ \subseteq\ \Ball(\rho,\,R+18\tau)\cap U$.
\end{lemma}
The second term holds for every representative $\rho$ and every $R\ge0$.

\begin{proof}
For the first inclusion, let $x\in U$ with $d(o,x)\le r$ and let $\rho'$ be the representative of $x$'s cluster.
Since $z$ and $x$ lie within $6\tau$ of $\rho$ and $\rho'$ respectively,
\[
  d(\rho,\rho')\ \le\ d(\rho,z)+d(z,o)+d(o,x)+d(x,\rho')\ \le\ 6\tau+r+r+6\tau\ \le\ R+12\tau,
\]
so $X_{\rho'}\subseteq Y(\rho,R)$ and in particular $x\in Y(\rho,R)$.
Note that $o$ itself need not lie in $U$.

For the second, let $x\in X_{\rho'}\subseteq Y(\rho,R)$, so $d(\rho,\rho')\le R+12\tau$ and $d(\rho',x)\le6\tau$; hence $d(\rho,x)\le R+18\tau$.
\end{proof}

We now show that $\Peel$ is correct and bound its strong queries.

\begin{theorem}
\label{thm:fptmsr}
 Set $L=\frac{10}{(1/2 - \delta)^{2}} (\log n+k\log k)$.\footnote{The larger value of $L$ is because we will invoke \Cref{lem:median} an FPT number of times.}
With probability at least $1-n^{-\Omega(1)}$, $\Peel(X,k)$ returns a feasible MSR solution with at most $k$ balls and total cost at most $40\,\optR(X,k)$, using $f(k)\cdot O(k\log n\log\log n)$ strong queries where $f(k)=k! (\log k)^{O(k)}=2^{O(k\log k)}$.
\end{theorem}

\begin{proof}
Let $\mathcal E$ be the event that the guarantee of \Cref{lem:sketch} holds for every call to $\Sketch$ made anywhere in the recursion tree.
We verify $\Pr[\mathcal E]\ge1-n^{-\Omega(1)}$ at the end, and condition on $\mathcal E$ throughout.

\medskip\noindent\textbf{Feasibility and the ball budget.}
Every branch adds one ball and decrements $m$, and $Y(\rho,R)\supseteq X_\rho\ne\emptyset$, so the recursion terminates within $k$ levels.
On $\mathcal E$, \Cref{lem:sandwich} gives $Y(\rho,R)\subseteq\Ball(\rho,R+18\tau)$ in \emph{every} branch, so the ball added at a level covers everything that level removes; together with the base case, any candidate returned covers $U$.
A candidate produced after $\ell$ levels uses $\ell$ balls plus at most $m=k-\ell$ from the base case, so at most $k$ in total.

\medskip\noindent\textbf{Cost.}
We exhibit one branch of cost at most $40\,\optR(X,k)$; since the algorithm returns the cheapest feasible candidate, this suffices.

Let $U_\ell$ be the residual at level $\ell$, $k_\ell=k-\ell$, and $\Phi_\ell=\optR(U_\ell,k_\ell)$.
We maintain the invariant $\Phi_\ell<\infty$, which holds at $\ell=0$.
If $|U_\ell|\le k_\ell$ the base case fires at cost $0$ and we stop, so assume $|U_\ell|>k_\ell$.
 Fix an optimal solution $\{\Ball(o_i,r_i)\}_{i\le k_\ell}$ for $(U_\ell,k_\ell)$, discarding any ball that contains no point of $U_\ell$, and let $\Ball(o^\star,r_\ell)$ be a ball of largest radius $r_\ell=\max_ir_i$.

\emph{The guess grid contains a good $R$.}
By \eqref{eq:cen-le-rmax} and \Cref{lem:sketch}(i), $\tau\le2\,\optC(U_\ell,k_\ell)\le2r_\ell$, so $r_\ell\ge\tau/2$.
In the other direction, the sketch itself is an MSR solution for $(U_\ell,k_\ell)$ with at most $k_\ell$ balls of radius $6\tau$, whence $r_\ell\le\Phi_\ell\le6k_\ell\tau$.
Thus $2r_\ell\in[\tau,\,12k_\ell\tau]$; and the grid $\mathcal G$ contains some $R$ with $2r_\ell\le R\le4r_\ell$.

\emph{The intended branch.}
Pick any $z\in\Ball(o^\star,r_\ell)\cap U_\ell$, and let $\rho$ be the representative of the cluster containing $z$; take the branch $(\rho,R)$.
By \Cref{lem:sandwich} with $r=r_\ell$,
\[
  \Ball(o^\star,r_\ell)\cap U_\ell\ \subseteq\ Y(\rho,R)\ \subseteq\
  \Ball(\rho,\,R+18\tau)\cap U_\ell .
\]
The left inclusion says $U_{\ell+1}=U_\ell\setminus Y(\rho,R)$ is covered by the remaining $k_\ell-1$ optimal balls, whose centers in $X$ are unaffected by the deletion, so $\Phi_{\ell+1}\le\Phi_\ell-r_\ell<\infty$ and the invariant is maintained.
The right inclusion says the ball we add has radius
\[
  R+18\tau\ \le\ 4r_\ell+18\cdot2r_\ell\ =\ 40\,r_\ell .
\]
Summing over levels and telescoping,
\[
  \sum_{\ell}40\,r_\ell\ \le\ 40\sum_\ell\bigl(\Phi_\ell-\Phi_{\ell+1}\bigr)\ \le\
      40\,\Phi_0\ =\ 40\,\optR(X,k).
\]

\medskip\noindent\textbf{Branching factor and queries.}
At a level with budget $m$ there are at most $m$ representatives and $|\mathcal G|=O(\log m)$ guesses, so the number of recursion-tree nodes is at most
\[
  \prod_{m=1}^{k}O(m\log m)\ =\ k!\,(\log k)^{O(k)}\ =:\ f(k)\ =\ 2^{O(k\log k)} .
\]
Each node calls $\Sketch$ once, costing $O(kL\log\log n)$ strong queries (\Cref{lem:sketch}).
Since $L=\Theta(\log n+k\log k)$,
\[
  f(k)\cdot O\bigl(k(\log n+k\log k)\log\log n\bigr)
  \ =\ f(k)\cdot O(k\log n\log\log n),
\]
after absorbing the factor $k^2\log k$ into $f(k)=2^{O(k\log k)}$.
Weak queries are counted by distinct pairs, of which there are trivially at most $\binom n2$; all remaining work is polynomial per node.

\medskip\noindent\textbf{The event $\mathcal E$.}
Each call to $\Sketch$ evaluates at most $O(nk\log\log n)$ median tests, and there are at most $f(k)$ calls, so at most $f(k)\cdot O(nk\log\log n)$ tests overall.
As a point is a witness in at most one execution (\Cref{sec:fpt-alg}), no test uses a weak answer that an earlier test has seen.
So by \Cref{lem:median} each test fails with probability at most $e^{-2(1/2-\delta)^2L}$, and a union bound over all tests gives
\[
  \Pr[\overline{\mathcal E}]\ \le\ f(k)\cdot O(nk\log\log n)\cdot e^{-2(1/2-\delta)^2L}
  \ \le\ n^{-\Omega(1)}
\]
once $L=\Omega(\log n+\log f(k))=\Omega(\log n+k\log k)$, which is exactly our choice of $L$.
\end{proof}

\subsection*{Discussion}
\label{sec:discussion}

For $k$-center, our bound of $O(k\log n\log\log n)$ strong point queries is within $O(\log n\log\log n)$ of the $\Omega(k)$ lower bound of Bateni et al.~\cite{bateni2024metric}, and the main open question is whether the $\log n$ factor is necessary.
\Cref{thm:rowlocal} shows that it is necessary for row-local algorithms, but we do not know whether every algorithm needs $\omega(k)$ strong point queries, that is, whether the $\Theta(\log n)$ gap is an artifact of row-locality.
Answering this needs a different instance: on ours, an algorithm that compares unqueried points with each other recovers every cluster with no strong queries (\Cref{prop:barrier}).

For MSR, the sketch of \Cref{lem:sketch} costs only $O(k\log n\log\log n)$ strong queries, so the factor $f(k)$ in \Cref{thm:fptmsr} comes entirely from branching; whether a constant factor is achievable with $\poly(k)\cdot\polylog(n)$ strong point queries is open.

\paragraph*{Declaration on the use of AI.}
AI-based tools were used in the preparation of this manuscript as follows.
The TikZ code for the figures was produced with AI assistance, and AI tools were used to help write and edit some of the prose.
The hard instance for the row-local lower bound was developed through interaction with GPT 5.6 sol; the class of row-local algorithms was the authors' idea, and the current simplified construction and its presentation are the authors' own.
All other research ideas and technical content are due solely to the authors.
The authors take full responsibility for all content.

\section*{Acknowledgments}
A.V. is supported in part by NSF CAREER award CAIF-2638918.
Z.F. is funded by an NSERC Discovery Grant.

\end{document}